\documentclass[12pt]{article}
\usepackage{latexsym,epsfig,graphicx,epstopdf,amsmath,amssymb,amscd,undertilde,multirow,chicago,paralist,dsfont,url, hyperref,subcaption, cite, ragged2e, booktabs}
\usepackage{amsfonts}
\usepackage{mathtools}
\usepackage[titletoc]{appendix}
\usepackage{rotating}

\newcommand{\thetavec}{{\boldsymbol{\theta}}}

\newcommand{\wvec}{{\boldsymbol{w}}}
\newcommand{\yvec}{{\boldsymbol{y}}}

\newcommand{\onevec}{{\boldsymbol{1}}}

\newcommand{\betavec}{{\boldsymbol{\beta}}}

\newcommand{\alphavec}{{\boldsymbol{\alpha}}}

\newcommand{\Xvec}{\boldsymbol{X}}

\newcommand{\xvec}{\boldsymbol{x}}

\newcommand{\Xmat}{X}
\newcommand{\Wmat}{W}
\newcommand{\Kmat}{K}
\newcommand{\Imat}{I}

\newcommand{\Xtensor}{\boldsymbol{\Xmat}}
\newcommand{\Itensor}{\boldsymbol{\Imat}}
\newcommand{\Ktensor}{\boldsymbol{\Kmat}}

\newcommand{\TP}{{\text{TP}}}
\newcommand{\TN}{{\text{TN}}}
\newcommand{\FP}{{\text{FP}}}
\newcommand{\FN}{{\text{FN}}}
\newcommand{\Accuracy}{{\text{Acc}}}
\newcommand{\Precision}{{\text{Precis}}}
\newcommand{\MacroAveragePrecision}{{\text{MacroAP}}}
\newcommand{\MacroAverageRecall}{{\text{MacroAR}}}
\newcommand{\Recall}{{\text{Recall}}}
\newcommand{\BalancedAccuracy}{{\text{BalAcc}}}
\newcommand{\F}{{\text{F}}}
\newcommand{\degreeO}{{^{\circ}}}
\newcommand{\MacroF}{{\text{MacroF}}}
\newcommand{\WeightedF}{{\text{WeightedF}}}
\newcommand{\total}{{\text{Total}}}
\newcommand{\MCC}{{\text{MCC}}}

\usepackage[american]{babel}

\begin{document}


\title{A Comprehensive Case Study on the Performance of Machine Learning Methods on the Classification of Solar Panel Electroluminescence Images}


\author{
Xinyi Song$^1$, Kennedy Odongo$^2$, Francis G. Pascual$^3$, and Yili Hong$^1$\\[1.5ex]
{\small $^1$Department of Statistics, Virginia Tech, Blacksburg, VA 24061}\\
{\small $^2$School of Business, Hamline University, St Paul, MN 55104}\\
{\small $^3$Department of Mathematics and Statistics, Washington State University, Pullman, WA 99164}
}


\date{}

\maketitle
\begin{abstract}
Photovoltaics (PV) are widely used to harvest solar energy, an important form of renewable energy. Photovoltaic arrays consist of multiple solar panels constructed from solar cells. Solar cells in the field are vulnerable to various defects, and electroluminescence (EL) imaging provides effective and non-destructive diagnostics to detect those defects. We use multiple traditional machine learning and modern deep learning models to classify EL solar cell images into different functional/defective categories. Because of the asymmetry in the number of functional vs. defective cells, an imbalanced label problem arises in the EL image data. The current literature lacks insights on which methods and metrics to use for model training and prediction. In this paper, we comprehensively compare different machine learning and deep learning methods under different performance metrics on the classification of solar cell EL images from monocrystalline and polycrystalline modules. We provide a comprehensive discussion on different metrics. Our results provide insights and guidelines for practitioners in selecting prediction methods and performance metrics.

\end{abstract}

\newpage

\section{Problem Description}
\subsection{The Problem}
Photovoltaics (PV) is an important form of renewable technology that generates electricity from solar radiation. In PV electricity generation, a PV array consists of multiple solar panels, and a solar panel contains multiple modules that are assemblies of PV cells. Monocrystalline silicon and polycrystalline silicon are the two main types of materials that are used to build PV cells (\shortciteNP{PVshareSM2016}). PV cells (also known as solar cells) in the field are vulnerable to various types of defects that eventually affect the functionality of the PV cells. For example, micro-cracks on the PV cell can affect the efficiency of solar panels.  Some hot spots on the panel can also affect the overall performance. Thus, identifying defects to ensure the reliability of PV cells is essential for performance management in the PV power industry.

\begin{figure}
\begin{center}
\begin{tabular}{ccccc}
 \parbox[t]{2mm}{\rotatebox[origin=l]{90}{\quad Monocrystalline}} &
\includegraphics[width = 0.21\textwidth]{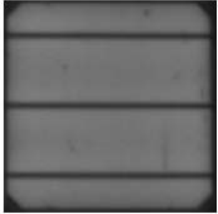}&
\includegraphics[width = 0.21\textwidth]{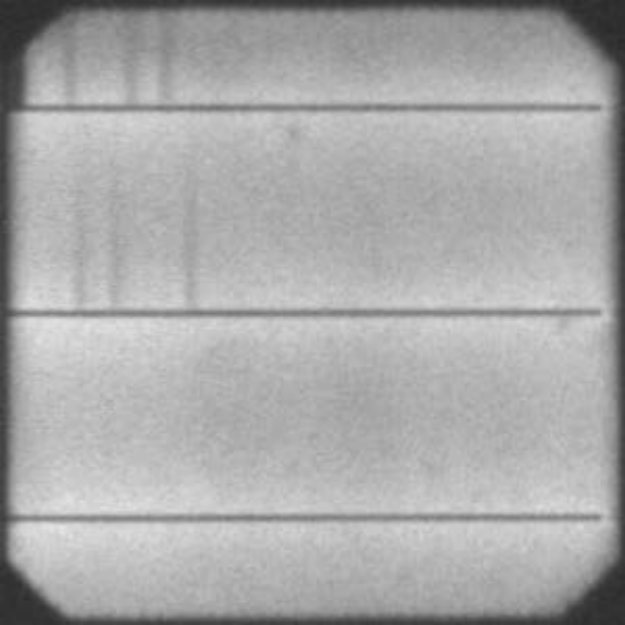}&
\includegraphics[width = 0.21\textwidth]{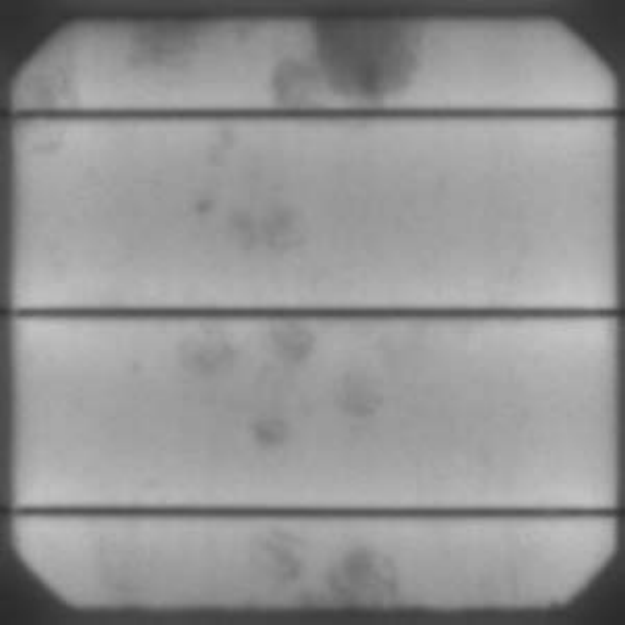}&
\includegraphics[width = 0.21\textwidth]{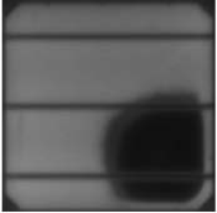}\\
 & (a) Functional & (b) Mildly D & (c) Moderately D & (d) Severely D \\[1ex]
\parbox[t]{2mm}{\rotatebox[origin=l]{90}{\quad Polycrystalline}} &
\includegraphics[width = 0.21\textwidth]{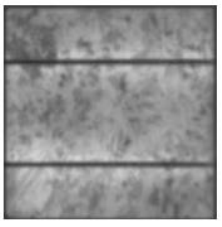}&
\includegraphics[width = 0.21\textwidth]{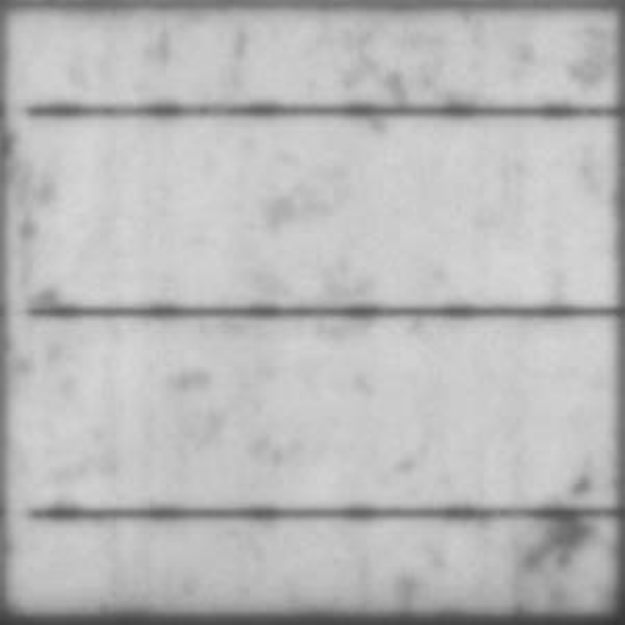}&
\includegraphics[width = 0.21\textwidth]{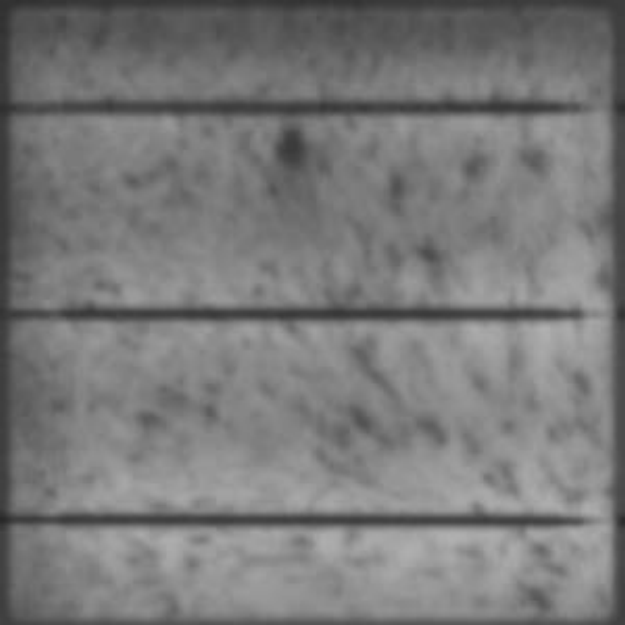}&
\includegraphics[width = 0.21\textwidth]{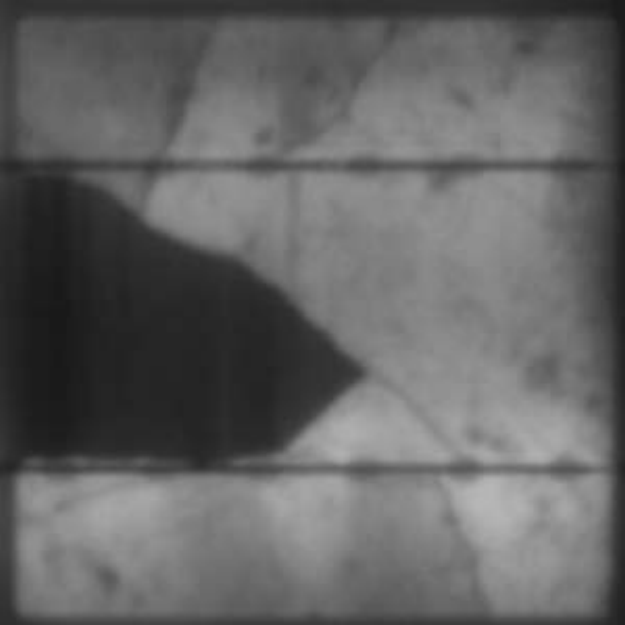}\\
 & (e) Functional & (f) Mildly D & (g) Moderately D & (h) Severely D \\
\end{tabular}
\caption{Examples of images with different severity of defectiveness in monocrystalline and polycrystalline PV cells. Here, the ``D'' in the label means ``defective''.}\label{fig:example.EL.image}
\end{center}
\end{figure}

To identify defects in PV cells, various methods have been used, which include manual disassembly and inspection, thermography imaging, and electroluminescence (EL) imaging. Among these inspection techniques, EL imaging is an effective and non-destructive detection method. In EL imaging, a direct current is applied to PV modules, and the emitted light from this process has a peak wavelength of 1150 nm (\shortciteNP{wave11502005}). Although we defer the details of the data to Section~\ref{sec:data.collection.prep}, Figure~\ref{fig:example.EL.image} shows examples of EL images with different conditions from both monocrystalline and polycrystalline PV cells.

Once EL images are obtained, technicians do a visual inspection of EL images to look for micro-cracks on the PV cells. This requires a large number of trained specialists, time, and budget (\shortciteNP{su12166416}). Alternatively, various machine learning and deep learning (ML/DL) methods can be used to automatically detect defects via EL images. Conceptually, an ML/DL algorithm classifies an EL image as defective, non-defective, or into other intermediate categories.

Due to variations in photovoltaic (PV) materials, diverse data inclusion criteria, and imbalanced data problems in different ML/DL approaches, it remains uncertain which method performs the best in EL image classification.
Additionally, it is important to acknowledge that different metrics can be utilized, and there are multiple approaches to determine the ``best'' method. Thus, in this case study paper, we aim to conduct a comprehensive comparison of different ML/DL methods under different evaluation metrics on the classification of solar cell EL images in both monocrystalline and polycrystalline PV modules with four defectiveness categories. We believe that insights from such a study will be valuable for practitioners in the field of PV reliability.

\subsection{Related Literature and Contributions of This Work} \label{sec:relatedwork}

A good amount of research has been devoted to automatically detecting defects in PV modules from EL images using ML/DL models. In the area of extracting features from EL images, \shortciteN{ASCDRA} used principal component analysis and Log-Gabor filters to extract features from EL images. \shortciteN{SergiuDeitschEL2019} discussed multiple feature extraction methods, such as scale-invariant feature transform, histograms of oriented gradients and speeded-up-robust features. On the other hand, convolutional neural network (CNN) has been widely used in image analysis, pattern recognition, and object detection without the need for feature engineering (e.g., \shortciteNP{Goodfellow2016}).  In addition to good prediction performance, CNNs can also be used for feature extraction. For instance, \shortciteN{DEMIRCI2021114810} proposed a deep feature-based method that uses DL to extract features from EL images. \shortciteN{Sovetkin2020EncoderDecoderSS} used encoder-decoder neural networks to conduct semantic segmentation of EL images.

Regarding DL models, variants of CNN models include Visual Geometry Group (VGG) in \citeN{vgg16}, Residual Neural Network (ResNet) in \shortciteN{resnet2016he}, GoogLeNet in \shortciteN{7298594}, and AlexNet in \shortciteN{Alexnet2017}. These models have been applied to detect defects in PV modules from EL images. There are also studies focused on hybrid methods. For example, \citeN{MRUDefectinspec2020} considered a hybrid loss that incorporates both focal and dice loss with multi-attention U-net. \shortciteN{DenoisingCNN} proposed a novel framework that creates the serial connection of de-noising CNN and ResNet-50 model. Because sufficient sample size is required when training complex DL models, many data augmentation methods were proposed for EL image data analysis (e.g., \shortciteNP{WQ994713}, and \shortciteNP{TANG2020453}).

In summary, various approaches have been developed to detect defects in PV modules, each utilizing EL image data collected under different procedures, resulting in different data distributions.  Also, it is worth noting that many studies primarily evaluate model performance based on accuracy, which can be less informative in the context of imbalanced problem settings. Therefore, in our study we aim to conduct a comprehensive study that evaluates the performance of ML/DL models for defectiveness detection using multiple metrics, based on published EL image datasets (\shortciteNP{Buerhop2018}).

\begin{figure}
	\begin{center}
		\includegraphics[width=1.0\textwidth]{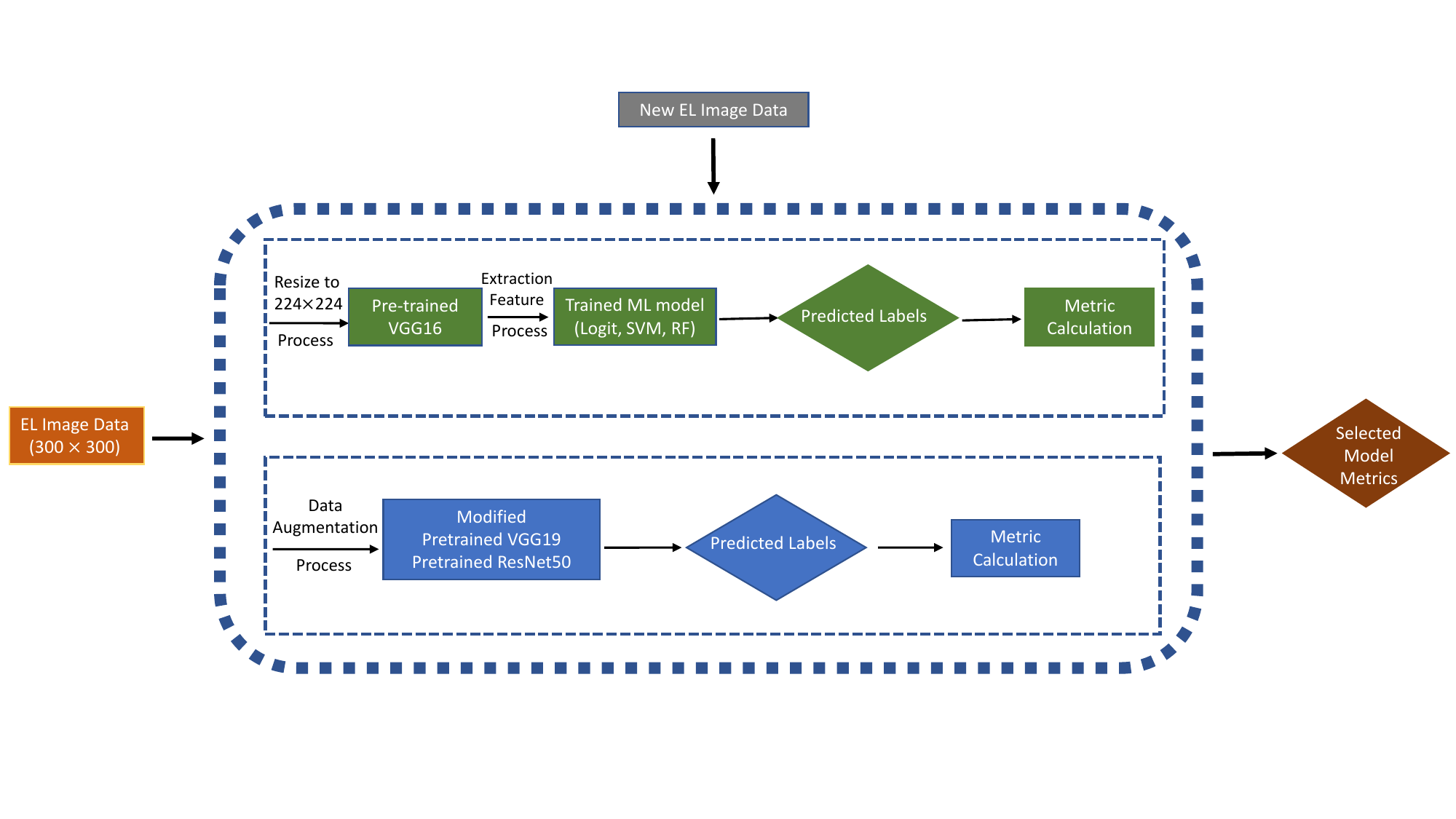}
		\caption{Flowchart that illustrates how the ML and DL methods are implemented in the performance study.}
		\label{flowchart}
	\end{center}
\end{figure}

The main contribution of this work is summarized as follows. Unlike existing literature, we consider both monocrystalline and polycrystalline photovoltaic modules in our accuracy and metric studies. We consider four class labels as described in Figure~\ref{fig:example.EL.image}, which leads to imbalanced classification problems. We conduct extensive comparisons among various ML/DL models using different performance metrics. Additionally, we provide an in-depth discussion of these metrics and provide practical guidelines for practitioners in selecting suitable prediction methods and performance metrics. To help readers see the overall picture of this paper, a step-by-step flowchart for all the included ML and DL methods is shown in Figure~\ref{flowchart}, and the details of those elements in the figure will be discussed in the following sections.

\subsection{Overview}
The rest of this paper is organized as follows. Section~\ref{sec:data.collection.prep} presents the PV EL data collection and data pre-processing. Section~\ref{sec:analysis.interpretation} gives a brief introduction to ML/DL models, metrics for comparisons, and data analysis results and interpretations.
Section~\ref{sec:conclusion.remark} contains some concluding remarks and some areas for future research.

\section{Data Collection and Preparation}\label{sec:data.collection.prep}
\subsection{Data Collection}
The PV EL dataset includes 2,624 solar cell images in total taken from 44 PV modules (\shortciteNP{Buerhop2018}), among which 18 are from monocrystalline modules and the rest are from polycrystalline modules. Both types of PV modules transform solar energy into electricity, but they are made from different components: a monocrystalline solar panel is composed of a single silicon crystal, while a polycrystalline solar panel is made up of multiple fused silicon pieces. Usually, polycrystalline panels are less efficient in terms of power output per square meter than monocrystalline panels but their manufacturing process is simpler and more cost effective.

There are four categories used to label the EL images in this paper: functional, mildly defective, moderately defective, and severely defective. A functional PV cell in an EL image usually displays a uniform brightness and consistent texture throughout. The image would show an even distribution of luminescence without any dark spots, lines, or irregular patterns that could indicate cell inefficiency or damage as shown in Figure~\ref{fig:example.EL.image} for both monocrystalline and polycrystalline PV modules. The functional cell operates at or near its expected efficiency (\shortciteNP{FailureCriterion2014}). When assessing defects in solar cells, experts not only categorize them as functional or defective but also assign a confidence level to their evaluations. They first determine whether the solar cell is functioning properly or displays any malfunctions or anomalies, and then they specify how confident they are in their assessments, providing a measure of certainty about their decisions. A weighting system is applied to determine the label for each image according to the result of the assessment and the level of belief.  In particular, if an EL image is assigned with a certainty weight of 33$\%$ (indicating a lower level of certainty) in the evaluation for defectiveness, the image is labeled as ``mildly defective''. Similarly, if an image is assigned with a certainty weight of 67$\%$ (reflecting a greater certainty in identifying defects), the image is categorized as ``moderately defective''. This labeling system takes into consideration both their functional status and certainty of evaluation, which brings more accurate and comprehensive defectiveness information.

\begin{figure}
\begin{center}
		\begin{tabular}{cc}
			\includegraphics[width=.48\textwidth]{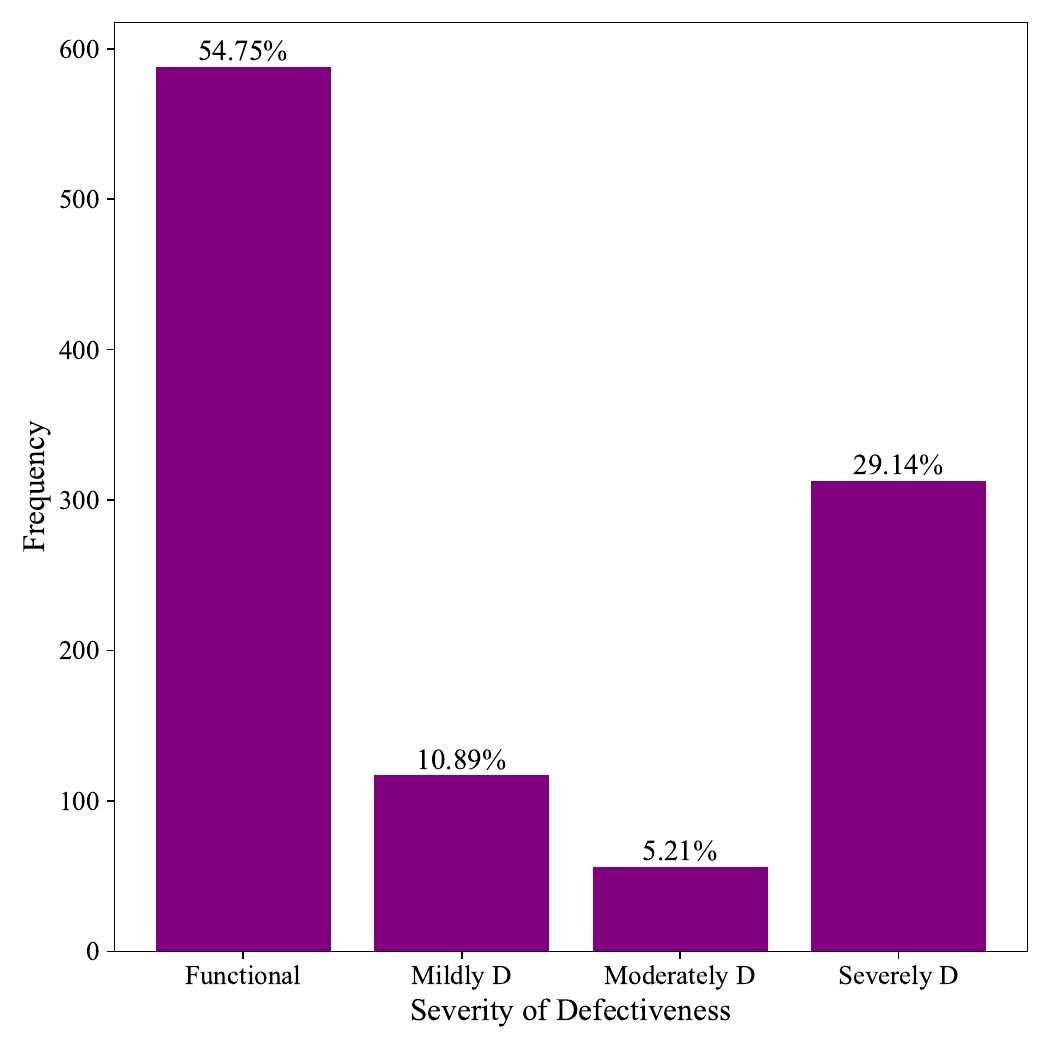} &
			\includegraphics[width=.48\textwidth]{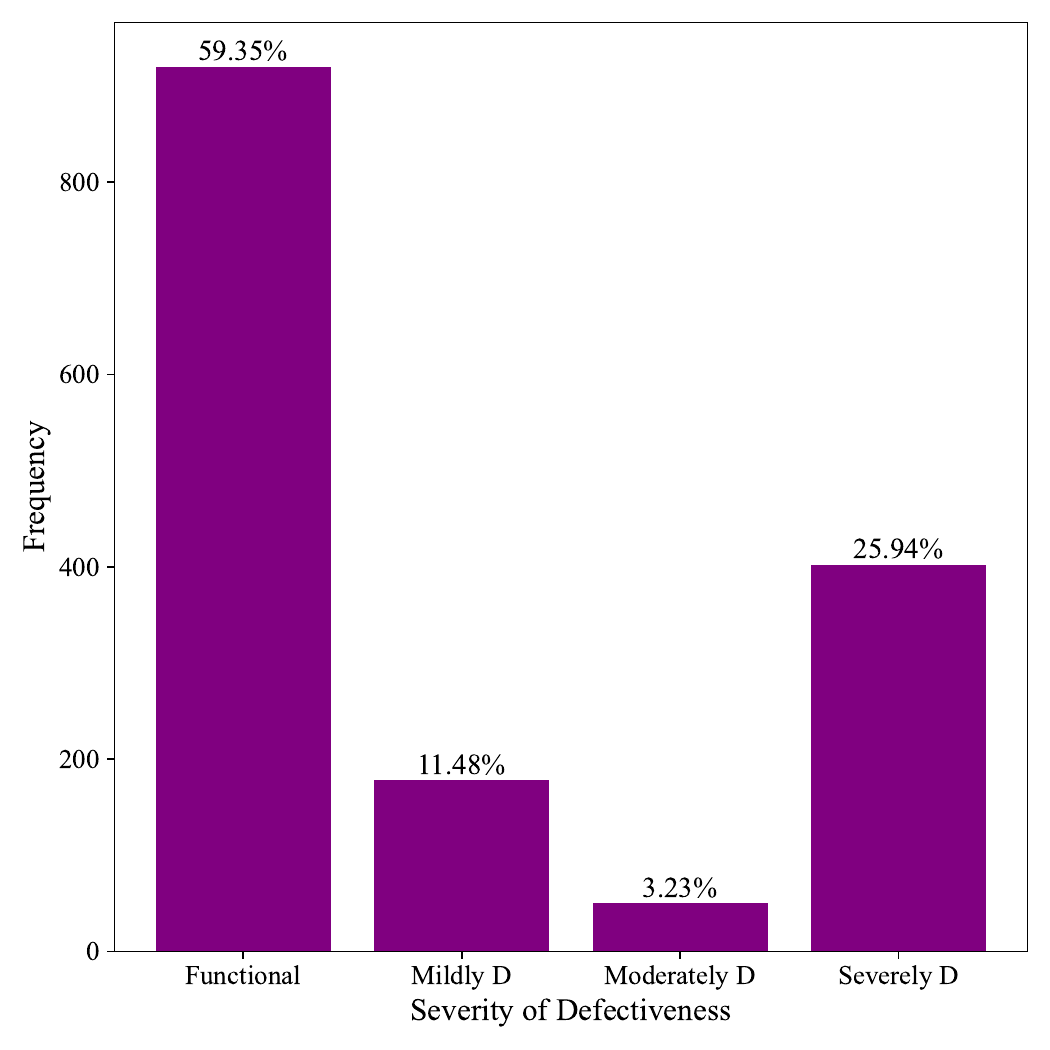}\\
			(a) Monocrystalline & (b) Polycrystalline \\
		\end{tabular}
		\caption{Bar plots of defectiveness counts in monocrystalline (a) and polycrystalline (b). Here the ``D'' in the $x$-axis label means ``defective''. One can see that an imbalanced problem exists in EL image data of both polycrystalline and monocrystalline.}
		\label{fig:hist_mono_poly}
	\end{center}
\end{figure}

Besides, the type of PV module (monocrystalline vs. polycrystalline) can influence the result of EL imaging analysis. Monocrystalline cells may allow for more straightforward identification of defects because of their uniform crystal structure. Cracks and defects disrupt the uniformity, making them more apparent in EL images, which can be seen in Figure \ref{fig:example.EL.image}. For image of ``severely defective'' class, it has a dark spot. However, polycrystalline cells require a more nuanced analysis due to the varied crystal structures. Defects might not be as immediately apparent and could be confused with the natural textures of the silicon grains (\shortciteNP{FailureCriterion2014}). Distinguishing between functional, mildly defective, moderately defective, and severely defective cells through EL images is difficult for non-experts, as the nuances may not be readily apparent to those untrained eyes. Therefore, we consider both monocrystalline and polycrystalline to better capture the properties of PV modules for severity of defectiveness identification. Table~\ref{tab:image.counts} summarizes the counts for the solar cell images. Figure~\ref{fig:hist_mono_poly} presents a bar plot of the counts of each defect category by PV module type. Stratified sampling is used to divide the images, maintaining the original population distribution across both PV modules. Specifically, a random selection process ensures that 80$\%$ of the images from each module class are designated for training, with the remaining 20$\%$ set aside for testing purposes. By adopting this method, the sample proportions of PV module types of the original population's distribution are preserved, allowing for a more representative training and testing data split for the EL images.

\begin{table}
\begin{center}
\caption{Summary of image counts. Here, ``Poly'' means polycrystalline, ``Mono'' means monocrystalline, and ``D'' means defective. ``Original'' denotes the EL image dataset, ``Augmented'' refers to the EL image data with artificially increased size by applying various transformations and modifications as illustrated in Section \ref{sec:feature.extraction}.}\label{tab:image.counts}.
\begin{tabular}{ccccccc}
\hline\hline
Sources & Types & Functional & Mildly D & Moderately D & Severely D & Total\\\hline
\multirow{2}{*}{Original} & Poly & 920 & 178 & 50 & 402 & 1{,}550 \\ \cline{3-7}
 & Mono & 588 & 117 & 56 & 313 & 1{,}074 \\\hline
\multirow{4}{*}{Augmented} & \multirow{2}{*}{Poly} & \multicolumn{4}{c}{Train Set Size: 9{,}300} & Test \\\cline{3-7}
 &  & 5{,}520 & 1{,}068 & 300 & 2{,}412 & 1{,}860 \\\cline{2-7}
 &\multirow{2}{*}{Mono} & \multicolumn{4}{c}{Train Set Size: 6{,}444} & Test \\\cline{3-7}
 &  & 3{,}528 & 702 & 336 & 1{,}878 & 1{,}289 \\\hline\hline
\end{tabular}
\end{center}
\end{table}

\subsection{Feature Extraction and Data Augmentation}\label{sec:feature.extraction}
\begin{figure}
	\begin{center}
		\includegraphics[width = 0.9\textwidth]{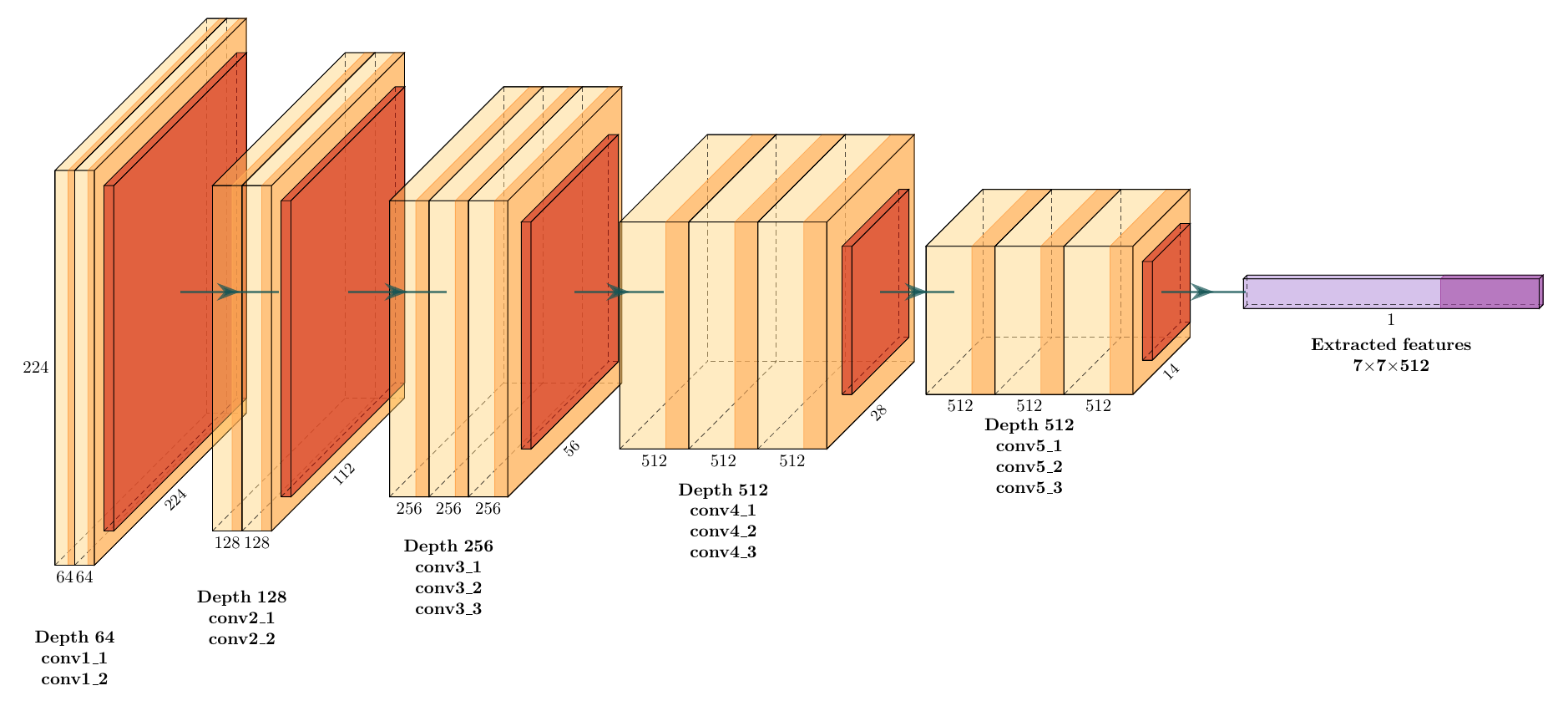}
		\caption{Visualization of the VGG-16 architecture for feature extraction, drawn by using the neural network visualization tool in Iqbal (2018).}
		\label{fig:vgg16}
	\end{center}
\end{figure}

When considering ML models discussed later in Section~\ref{sec:trad.machine.learning.models}, the input variable needs to be a vector, representing the features extracted from the images. In this paper, the pre-trained VGG-16 network (\shortciteNP{vgg16}) is employed for feature extraction which consists of two main parts: the feature extraction part, also known as the convolutional base as shown in Figure~\ref{fig:vgg16} (drawn by using the neural network visualization tool in \citeNP{haris}), and the classification part involving fully connected layers. The classification part is removed because it is responsible for final classification and only the convolution base is kept for feature extraction (\shortciteNP{AlMalla2022PretrainedCA}).  This pre-trained convolutional base expects an input image of size $224\times224$ and produces an output after the last max pooling layer with a dimension of $7\times7\times512$. The extracted features capture the hierarchical representations learned by the convolutional layers, providing a more comprehensive and informative representation of the input images. By utilizing feature engineering instead of flattening the whole original image tensor to a vector, computational resources can be conserved without compromising accuracy. Here, a tensor is a generalized version of matrix at a higher dimension.

In the pre-trained VGG-16 architecture, the convolutional base is loaded with pre-trained weights based on the ImageNet dataset (\shortciteNP{ImageNet2009}). To align with the requirements of convolutional base in pre-trained VGG-16, the EL image is resized from its original size of $300\times300$ to $224\times224$. Furthermore, the images are converted from grayscale (with a single channel) to RGB (with three channels) to match the required input format of VGG-16. The resulting tensor from the convolutional base has a dimension of (7, 7, 512). To use this as input for ML models, the tensor is flattened into a vector of length $7\times7\times 512 = 25{,}088$. This flattened vector denoted as $\xvec$, can then be utilized as the input feature vector in ML models.

\begin{figure}
\begin{center}
\begin{tabular}{ccccc}
\includegraphics[width = 0.17\textwidth]{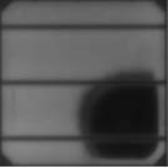}&
\includegraphics[width = 0.17\textwidth]{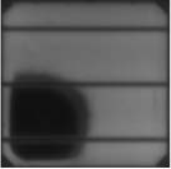}&
\includegraphics[width = 0.17\textwidth]{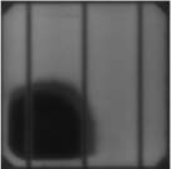}&
\includegraphics[width = 0.17\textwidth]{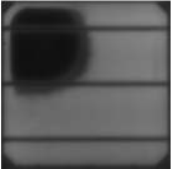}&
\includegraphics[width = 0.17\textwidth]{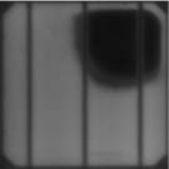}\\
 (a) Original & (b) Flip horizontally & (c) Rotate $90\degreeO$ & (d) Rotate $180\degreeO$ & (e)
 Rotate $270\degreeO$
\end{tabular}
\caption{Illustration of image augmentation.}\label{fig:image_aug}
\end{center}
\end{figure}

For DL models we consider later in Section~\ref{sec:deep.learning.models}, the preprocessing procedure includes image normalization and augmentation. Image normalization entails standardizing image data by adjusting pixel values to a uniform range and distribution, thereby facilitating more efficient model convergence. Here the EL images are normalized according to the mean and standard deviation of ImageNet for consistency and transferability of pre-trained weights. Also, data augmentation is applied to bring more diversity, which helps learn more general patterns of images for prediction and avoid overfitting. By horizontally and vertically flipping, randomly rotating 90, 180 and 270 degrees for the training dataset, minor changes are added to the EL image dataset to extract more information during sample processing as shown in Figure~\ref{fig:image_aug}. Table~\ref{tab:image.counts} summarizes the image counts after data augmentation.

\section{Analysis and Interpretation}\label{sec:analysis.interpretation}
In this section, we introduce various ML/DL models and several metrics for performance comparisons. Finally, we analyze the EL image data.

\subsection{Traditional Machine Learning Models}\label{sec:trad.machine.learning.models}
For ML methods, three emblematic classification models are chosen: the multi-class logistic regression model, support vector machine (SVM), and random forest (RF). The multi-class logistic regression model stands out for its simplicity, efficiency, and interpretability. SVM excels in high-dimensional spaces, especially when the number of features is much larger than that of samples. Random forest is an ensemble model that controls overfitting and enhances accuracy. Additionally, pre-trained VGG16 model is utilized for feature extraction due to its proven efficacy. The resulting output feature vector from pre-trained VGG16 model is represented as $\xvec_{i}$, which is a $d \times 1$ vector, with $d$ indicating the number of features. Using the features extracted by VGG-16 architecture as described in Section~\ref{sec:feature.extraction}, we have  $d$ = 25{,}088 (i.e., 7$\times$7$\times$512) for the solar panel EL image data. Therefore, the input variables in ML models are $\xvec_i$ for $i=1, \dots, N$ with $N$ observations, and the output variable is $y_i$ that takes values in $\{1 ,\dots, J\}$, where $J$ is the number of categories.

\subsubsection{Logistic Regression}
Logistic regression models the relationship between the categorical response variable and predictors. Multinomial logistic regression is used for the solar panel EL data because there are four categories of defectiveness for the EL images (e.g., \citeNP{GLMMJ1997}). Let $\pi_{ij}=\Pr(y_i=j)$ be the probability that observation $i$ belongs to category $j$. Note that $\sum_{j=1}^J\pi_{ij}=1$ for each observation $i$.
The multinomial logistic regression comes with the assumption that the log odds follows a linear model. With class $J$ as the baseline, the models for the first $J-1$ classes are,
\begin{align*}
\eta_{ij}=\log \left(\frac{\pi_{ij}}{\pi_{iJ}}\right) = \alpha_j+\xvec_i'\betavec_j, ~ j=1,\dots, J-1.
\end{align*}
Here, $\alpha_j$ is a constant (i.e., the model intercept), and $\betavec_j$ is the coefficient vector for $j=1,2,\ldots,J-1$. To fit this model, we need to maximize the penalized log-likelihood function,
\begin{equation*}
\l(\thetavec) = \sum_{i=1}^N \sum_{j = 1}^{J}y_{ij}\log(\pi_{ij}) = \sum_{i = 1}^N \left\{\sum_{j=1}^{J-1}y_{ij} \eta_{ij} - \log[1 + \sum_{j=1}^{J-1}\exp(\eta_{ij})]\right\} + \lambda \sum_{j = 1}^{J-1} \Vert \betavec_{j}  \Vert,
\end{equation*}
where $y_{ij}=\onevec(y_i=j)$, $\onevec(\cdot)$ is the indicator function, and $\thetavec= (\alphavec', \betavec')'$. Note that, in our setting, the input vector $\xvec_i$ is of high dimension as compared with $\thetavec$. That is, $d \gg N$, and thus the $L_2$ regularization is applied. After parameters are estimated, the probability of predicting observation with $\xvec_i$ as the $j$th class is computed as,
\begin{align*}
	\pi_{ij} = \frac{\exp(\alpha_j + \xvec_i'\betavec_j)}{1 + \sum_{j=1}^{J-1}\exp(\alpha_j + \xvec_i'\betavec_j)}, ~j = 1\cdots J-1, \text{ and }\,\,
	\pi_{iJ} = \frac{1}{1 + \sum_{j=1}^{J-1}\exp(\alpha_j + \xvec_i'\betavec_j)}.
\end{align*}
The final predicted label $y^{\ast}$ for observation with $\xvec_i^{\ast}$ is the class with the largest predicted probability. That is,
$
\widetilde{y}^{\ast}_i= \operatorname*{argmax}_{j} \{\pi_{ij}\}, \; i= 1,\ldots, N.
$

\subsubsection{Support Vector Machine}

SVM works by identifying a hyperplane that maximizes correct class separations, with each observation being classified according to its position relative to this hyperplane (e.g., \citeNP{SVM1995}). The multi-classification problem can be treated as a $J$-class problem, which creates $J$ SVM models (e.g., \citeNP{LINGRAS20073782}). For example, the $j$th SVM model is trained considering all of the training observations in the $j$th class as positive, and all other classes as negative.

Given observations $(\xvec_i, y_i), i=1, \dots, N$, the maximum-margin hyperplane in the $j$th model is $\phi(\xvec_i)'\wvec^{(j)} +  \xi^{(j)} = 0$, where $\wvec^{(j)}$ is the weight vector, $\xi^{(j)}$ the bias term, and $\phi(\xvec)$ is the kernel function. The model is trained by solving the following optimization problem,
\begin{equation*}
\min_{\wvec^{(j)},\, \xi^{(j)} \atop j = 1,\cdots, J}\frac{1}{2} \left\lVert \wvec^{(j)}\right\rVert^2 + \lambda \sum_{i=1}^N \epsilon_i^{(j)},
\end{equation*}
subject to
\begin{equation*}
\begin{cases}
\phi(\xvec_i)'\wvec^{(j)} + \xi^{(j)} \geq 1 - \epsilon_i^{(j)}, ~\text{if}~  y_i = j, \\
\phi(\xvec_i)'\wvec^{(j)} + \xi^{(j)} \leq -1 + \epsilon_i^{(j)}, ~ \text{if}~  y_i \neq j,
\end{cases}
 i = 1 ,\ldots, N.
\end{equation*}
Here, $\left\lVert \wvec^{(j)}\right\rVert^2$ is the normalized vector for weights and the dimension of $\wvec^{(j)}$ depends on how kernel function $\phi(\cdot)$ maps the feature $\xvec$. The slack variable $\epsilon_i^{(j)}\geq 0$ allows a single observation $(\xvec_i, y_i)$ to be on the wrong side of the margin of the hyperplane in the $j$th SVM model, and $\lambda$ is the penalty parameter.

Once parameters are estimated, the decision function for the observation ($\xvec_i^{\ast}, y_i^{\ast}$) in the $j$th SVM model is
$
f^{\ast}_j(\xvec_i^{\ast}) = \phi(\xvec_i^{\ast})' \wvec^{(j)}  + \xi^{(j)}.
$
The test observation is then predicted as belonging to class $j$ which has the largest decision function value. That is,
$
 \widetilde{y}_{i}^{\ast}= \operatorname*{argmax}_j\{f^{\ast}_j(\xvec_i^{\ast})\}, i=1, \dots, N.
$
In the EL image data, instead of training $J$ one-versus-others SVM models, we train one binary SVM classification model for each specific pair of classes, resulting in a total of $J(J-1)/2$ models. For the EL image data, we have $J=4$ and $6$ SVM models in total. Each binary classifier generates a prediction for its corresponding pair of classes and the final predicted class is the one that receives the largest number of ``votes''.

\subsubsection{Random Forest}

RF involves constructing a diverse ensemble of uncorrelated decision trees by randomizing both data sampling and feature selection process (e.g., \citeNP{breiman2001random}), which enables each tree to learn multiple patterns from different subsets of observations and features (e.g., \citeNP{Rokach2005}). The combination of these two randomizations could keep each tree diverse, reduce the chance of overfitting, and enhance the robustness of the model to the outliers and noisy data. Let $t$ be the index for a tree and $T$ be the total number of trees. With observations $(\xvec_i, y_i), i=1, \dots, N$, tree $t \in \{1,\ldots, T\}$ is formed using the following algorithm.

\begin{enumerate}[1)]
	\item Randomly select $N_b<N$ observations from the training set, allowing for replacement.
	\item Based on the $N_b$ observations selected at Step 1, generate an RF tree $t_{b}$ by recursively repeating the following steps for each terminal node of the tree until the minimum node size $n_{t_{\text{min}}}$ is reached:
	\begin{enumerate}
		\item Check whether the current node size is equal to or smaller than the $n_{t_{\text{min}}}$. If it is, stop the recursion,  consider the node as terminal, the creation of this tree is complete, and go to Step 3.
		\item Otherwise, perform the following steps:
		 \begin{enumerate}
		 	\item Randomly select $d_{t}$ features among the features in $\xvec_i$.
		 	\item Evaluate different splits using the selected $d_{t}$ feature subset and choose the best split based on the minimized value of Gini impurity. Gini impurity quantifies the probability of misclassifying an observation with ranges from 0 to 1, where 0 denotes the highest level of homogeneity and 1 indicates maximum impurity.
		 	\item Split the node based on the chosen split, creating two child nodes.
		 	\end{enumerate}
	 	\item Recursively repeat the Steps (a) and (b) above at each subsequent child node.
	 \end{enumerate}
 \item Repeat Steps 1 and 2 until an ensemble of the $T$ trees is created.
\end{enumerate}
Let $M_t(\xvec^{\ast})$ be the predicted class of $t$th tree for feature input $\xvec^{\ast}$. The final predicted class $y^{\ast}$ is,
\begin{equation*}
	 \widetilde{y}^{\ast} = \operatorname*{argmax}_{j} \sum_{t=1}^T \onevec(M_t(\xvec^{\ast}) = j),
\end{equation*}
where $\onevec(\cdot)$ is the indicator function.

\subsection{Deep Learning Models}\label{sec:deep.learning.models}
In the DL approach, we consider two deep transfer learning models based on VGG-19 and ResNet architecture. We first give an introduction to the CNN models (e.g., \shortciteNP{Goodfellow2016}). The input of a DL model in our case is an image that is represented by a tensor $\Xtensor_i$ for $i = 1 ,\ldots, N$, and the output is a prediction for the image class, $y_i$. The dimension of $\Xvec_i$ is $(n_{h}, n_{w}, n_{c})$, which is denoted by $\dim(\Xtensor) = (n_{h}, n_{w}, n_{c})$, representing the height, width and number of channels. For RGB images, $n_c = 3$ corresponds to the channels of red, green, and blue. A CNN model mainly consists of three types of layers: convolutional layers, pooling layers, and fully connected layers. A convolutional layer transforms images for feature extraction and is followed by a pooling layer for dimension reduction. Then the output from the pooling layer is flattened for the fully connected layer to calculate probability and make classification decisions.

In the convolution layer, a filter $\Ktensor$ (i.e., kernel) is used, which needs to be learned from the data. A filter has the same number of channels as the input, and the dimension is typically $\dim(\Ktensor) = (f, f, n_{c})$, where $f$ is the size of the filter. The output of a convolutional layer is a 2D matrix derived from the sum of the element-wise multiplication between the filter $\Ktensor$ and its covered cube of the input $\Itensor$. One starts with $\Itensor^{(0)}$ as the input image $\Xtensor$. Given the $q$th convolutional layer, the input is $\Itensor^{(q-1)}$ with dimension $(n_{h}^{(q-1)}, n_{w}^{(q-1)}, n_{c}^{(q-1)})$ and the output is $\Itensor^{(q)}$ with dimension $(n_{h}^{(q)}, n_{w}^{(q)}, n_{c}^{(q)})$.  Then, for all $r \in \{1,2, \ldots, n_c^{(q)}\}$,
\begin{equation*}
	\text{Conv}(\Ktensor, \Xtensor)_{w, v} = \sum_{a = 1}^{n_h} \sum_{b = 1}^{n_w}\sum_{c = 1}^{n_c} \Ktensor_{a, b, c}\Itensor_{w+a -1, v + b-1, c},
\end{equation*}
where $\text{Conv}$ stands for convolution, $\text{Conv}(\Ktensor, \Xtensor)_{w, v}$ gives the entry at the location $(w, v)$ that corresponds to the $w$th row and $v$th column of the output matrix calculated by element-wise multiplication, and $\Ktensor_{a, b, c}$ is the entry in location $(a, b, c)$ of tensor $\Ktensor$.

The pooling layer summarizes the features extracted by the convolution layer. Specifically, given the $q$th pooling layer, the input is $\Itensor^{(q-1)}$, where $q$ starts from 1. The pooling function is $\psi^{(q)}$ and the size of the filters is $f^{(q)}$. The output is $\Itensor^{(q)}$ with dimension $(n_{h}^{(q)}, n_{w}^{(q)}, n_{c}^{(q)})$. The computing is done as $\Itensor^{(q-1)}_{w, v, z} = \psi^{(q)}(\Itensor^{(q-1)}_{w+a-1, v+b-1, z})$, where $a \in \{1, 2, \ldots, f^{(q)}\}$ and $b \in \{1, 2, \ldots, f^{(q)}\}$. In the case of fully connected layers, the input tensor $\Itensor^{(q-1)}$ is transformed into a one-dimensional vector with a dimension of $(1, n_{h}^{(q-1)} \times n_{w}^{(q-1)} \times n_{c}^{(q-1)})$, making it suitable for processing in a fully connected layer. The output is $\Itensor_{p}^{(q)}$ =$\sigma^{(q)}(\sum_{l=1}^{n_{q-1}} \Wmat_{p,l}^{(q)}\Itensor^{(q-1)}+ \xi_{p}^{(q)})$ with weights $\Wmat_{p, l}^{(q)}$ and bias $\xi_{p}^{(q)}$. The activation function is $\sigma^{(q)}(\cdot)$. The final layer of fully connected layers utilizes the softmax activation function instead of ReLU to calculate the probability of each class.

For DL models, pre-trained architectures are employed considering the limited number of samples and computational resources. VGG-19, recognized for its profound architecture, facilitates the learning of feature hierarchies across several abstraction levels, significantly contributing to its superior performance in defect detection. However, it encounters a degradation problem from the increased complexity of optimizing such a deep network during training. To address this issue, the ResNet-50 architecture, which leverages ``residual blocks'' and skip connections to ensure that additional layers do not detrimentally affect model performance, is implemented in our study. Since DL models are capable of processing image data directly, 300$\times$300 images are fed into pretrained architectures including VGG-19 and ResNet-50. While they do not perform feature selection in the conventional sense, during the learning process, the DL models prioritize certain features over others via pooling layers and activation functions.

\subsubsection{VGG Neural Network}

\begin{figure}
	\begin{center}
		\includegraphics[width = 1\textwidth]{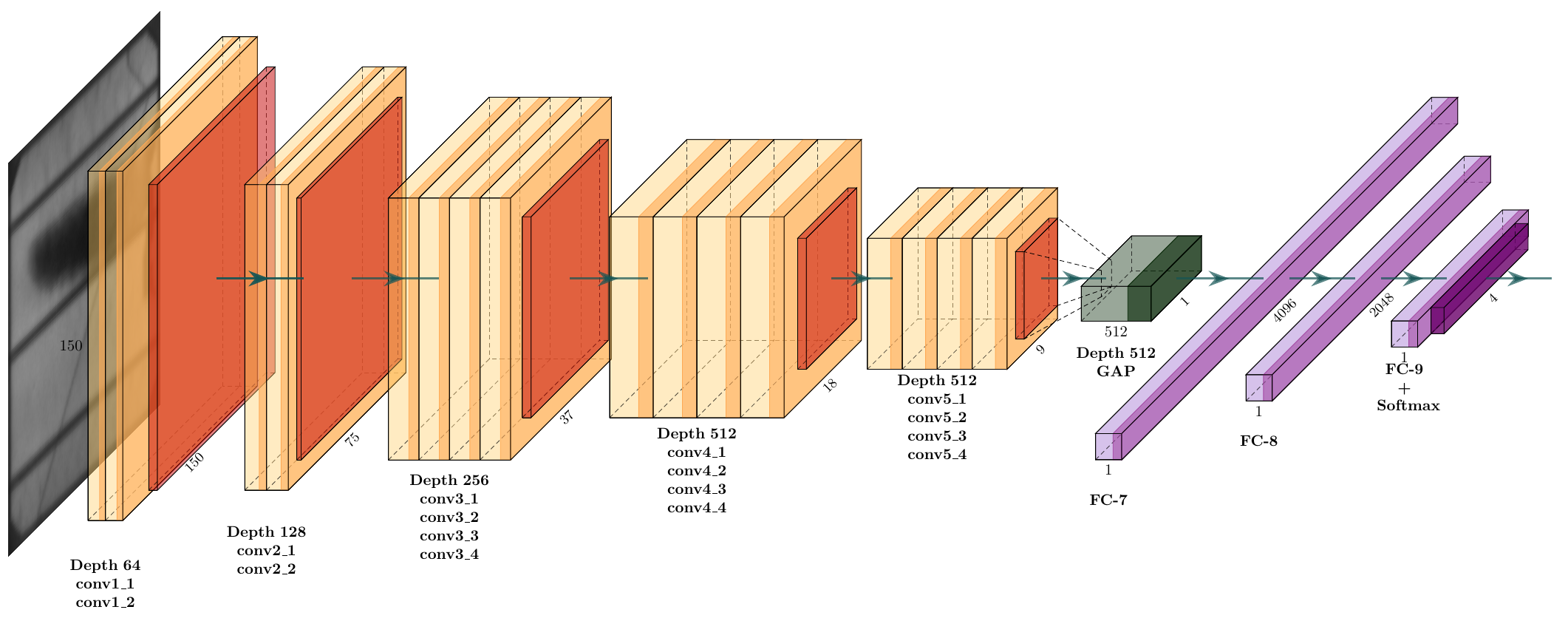}
		\caption{VGG-19 architecture, drawn by using the visualization tool in Iqbal (2018).}
		\label{fig:vgg19}
	\end{center}
\end{figure}

In this paper, we choose VGG-19 architecture (\citeNP{8474802}), which is a deep CNN with 19 convolution layers and 3 fully-connected layers, to classify the images pre-trained on the ImageNet (\shortciteNP{ImageNet2009}). To avoid further down-sampling of the solar cell images,  a global average pooling (GAP) layer is utilized to adapt the input tensor of the VGG-19 network (\shortciteNP{Deitsch2019}), which reduces the spatial dimensions while retaining the channel information. The final output of the modified VGG-19 neural network involves four neurons which correspond to the predicted probability for each class. The model is refined by categorical cross-entropy loss and its final decisions are made by softmax classifier. This architecture is refined according to the dimension of the EL image dataset as shown in Figure~\ref{fig:vgg19}.

The pre-trained VGG-19 model on EL image data is fine-tuned to identify the severity of defectiveness with two phases similar to \shortciteN{SergiuDeitschEL2019}. At the first phase among 50 epochs, the training procedure starts at fully connected layers with randomly initialized weights while freezing all parameters in the convolutional base. Stochastic gradient descent (SGD) optimizer is utilized at a learning rate of 0.01. Then at the second phase for another 50 epochs, all weight parameters are fine-tuned, still with SGD optimizer but at a smaller learning rate of 0.005.

\subsubsection{Residual Neural Network}

ResNet is the network with a skip connection (\shortciteNP{AD2018}), which is an identity connection that skips some layers to reduce degradation problems. With the help of the skip connection, the input from previous layers is passed to the next layer directly by matrix addition instead of multiplication as the usual convolution does. As a result, the skip connection does not increase the number of parameters.

\begin{figure}
	\begin{center}
		\includegraphics[width = .8\textwidth]{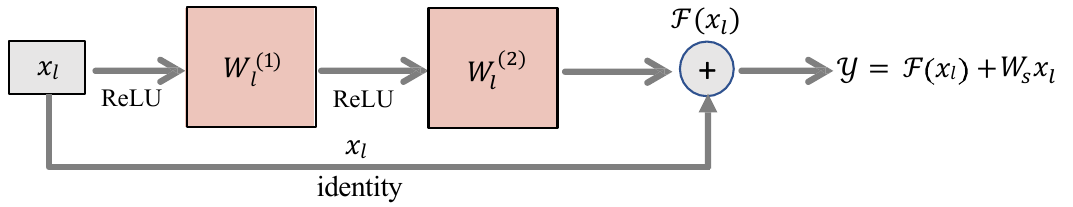}
		\caption{Visualization of the ResNet architecture.}
		\label{fig:resnet_example}
	\end{center}
\end{figure}

Given the $l$th residual block with the ReLU activation function and batch normalization, the input is denoted as $\xvec_{l}$, and the output is denoted as $\xvec_{l+1}$. Consider the following residual block as shown in Figure~\ref{fig:resnet_example},
\begin{equation}
	\yvec = \mathcal{F}(\xvec_{l}, \Wmat_{l}^{(k)}) +\xvec_{l} =\Wmat_{l}^{(2)}( \sigma^{(l)}(\Wmat_{l}^{(1)}\xvec_{l})) + \xvec_{l}.
	\label{eq:resnet}
\end{equation}
As seen in \eqref{eq:resnet}, the $\Wmat_{l}^{(k)}\xvec_{l}$ denotes the $k$th convolution process and $\sigma^{(l)}$ is the activation function in the $l$th module. Here, $\yvec$ is the vector for true outputs, and the residual is
$\mathcal{F}(\xvec_{l}, \Wmat_{l}^{(k)}) = \yvec - \xvec_{l}.$ This identity connection allows the residual block to focus on learning the residuals $\mathcal{F}(\xvec_l, \Wmat_l^{(k)})$.
Usually, if the dimensions of the input and the output are the same, identity skip connections are directly used. Here, ``identity'' refers to the fact that the connection simply carries the input data's identity from one layer to another without applying any transformation to it. However, sometimes, the dimensions of the input and the output are not the same when changing their channels. We consider two options. One option is that the skip connections still do identity mapping but pad extra zeros to accommodate for the increased dimensions without requiring additional parameters and computational cost. Another option is that the skip connections make a linear perceptron $\Wmat_s$ to match the required dimensions as shown in the following equations,
\begin{equation*}
	\yvec = \mathcal{F}(\xvec_{l}, \Wmat_{l}^{(k)}) +\Wmat_{s}\xvec_{l} =\Wmat_{l}^{(2)}( \sigma^{(l)}(\Wmat_{l}^{(1)}\xvec_{l})) + \Wmat_{s}\xvec_{l}, \quad \text{and}\quad
	\mathcal{F}(\xvec_{l}, \Wmat_{l}^{(k)}) = \yvec - \Wmat_{s}\xvec_{l}.
\end{equation*}

\begin{sidewaysfigure}

	\begin{center}
		\includegraphics[width = 1\textwidth]{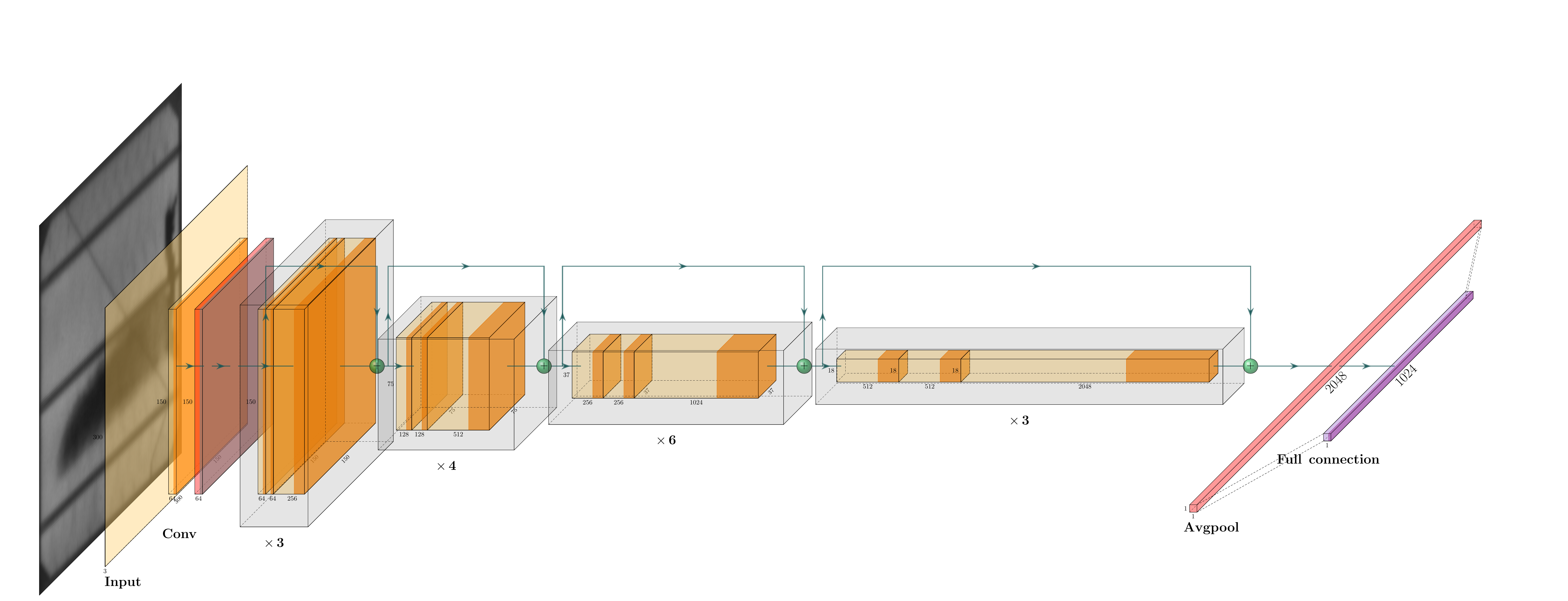}
		\caption{Visualization of the ResNet-50 neural network architecture, drawn by using the visualization tool in Iqbal (2018).}
		\label{fig:resnet50_arc}
	\end{center}
\end{sidewaysfigure}

The skip connection guarantees that the deeper layers would have at least as good performance as the shallower layers and provides the option of another shortcut path for the gradient to feed the network, which solves problems of gradient vanishing and degradation. ResNet-50 is utilized to identify the severity of defectiveness in our paper. It is an architecture with 50 layers of residual networks and comes with a special bottleneck design for the building block as shown in Figure~\ref{fig:resnet50_arc}. The bottleneck design entails several 1$\times$1 convolutions that reduce the number of parameters for higher model efficiency. Similar to the modification of VGG-19, GAP is added to avoid further down-sampling of the solar cell images and match the required dimension of the pre-trained network. The model is fine-tuned by an SGD optimizer at a learning rate of 0.005 with cross-entropy loss.

As one can see, pre-trained DL models are used in the current paper. The core idea of the pre-trained neural network is that the model is trained on a large dataset and then it transfers features learned to new datasets, which allows us to train the model with a small amount of data. If properly used, it can bring higher starting accuracy, faster convergence rate, and larger asymptotic accuracy.

\subsection{Evaluation Metrics}

\begin{figure}
	\begin{center}
		\includegraphics[width=.55\textwidth]{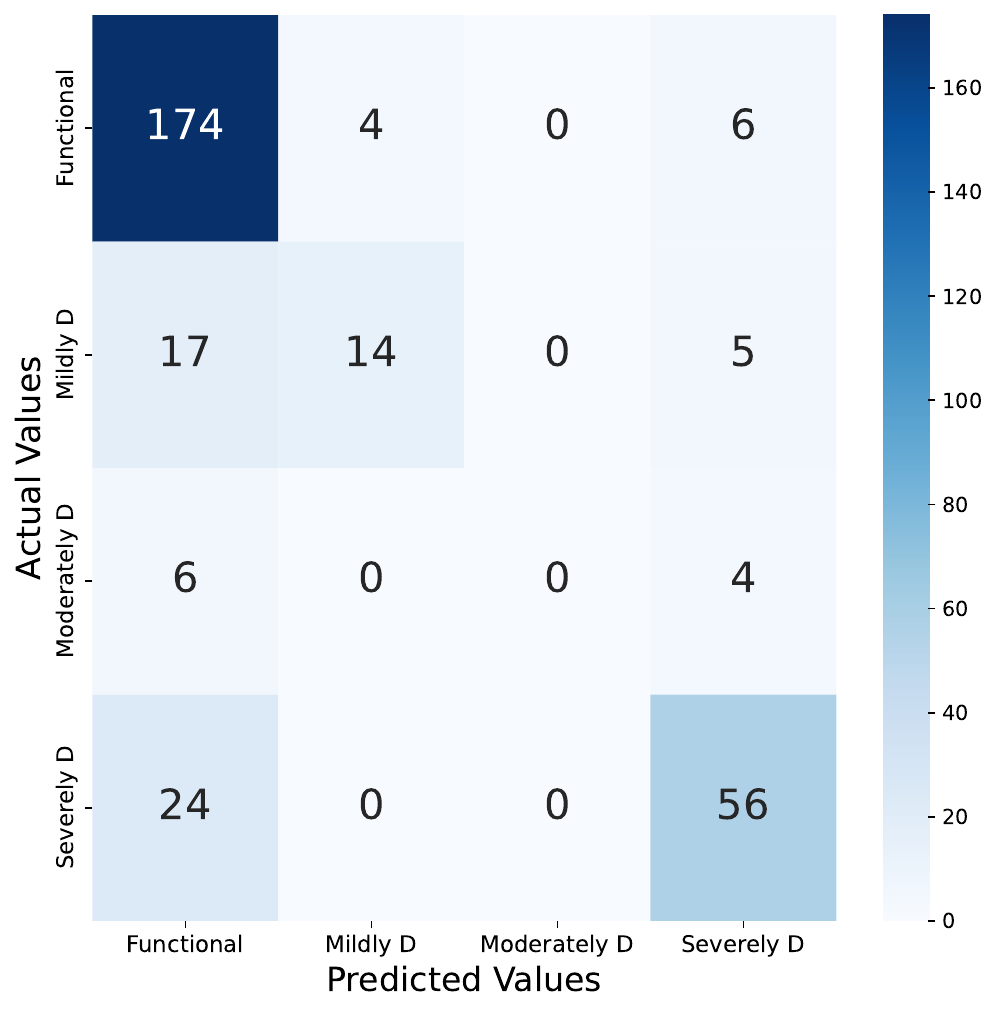}
		\caption{Illustration of a confusion matrix of an RF for classifying EL images of polycrystalline PV modules. Here, the rows represent ``predicted values'' and the columns denote ``actual values''.}
		\label{fig:cm_poly_rf_eg}
	\end{center}
\end{figure}

To better and comprehensively evaluate the model quality and performance, multiple metrics for multi-class cases are calculated and compared (e.g., \shortciteNP{Grandini2020MetricsFM}). A confusion matrix based on the RF classification of the PV modules test set is used to illustrate the calculations, as shown in Figure~\ref{fig:cm_poly_rf_eg}. We do not use the augmented data with ML models, and there are total of 310 observations of which 184, 36, 10, and 80 correspond to ``functional'', ``mildly defective'', ``moderately defective'', and ``severely defective'', respectively. True positive (TP) (true negative, TN) indicates an outcome where the model correctly classifies a positive (negative) instance as positive (negative). On the contrary, false positive (FP) (false negative, FN) occurs when the classifier incorrectly classifies a negative (positive) instance as positive (negative).

Specifically, among the 184 PV module observations identified as ``functional'', 174 are correctly classified as ``functional'', while 4 and 6 are incorrectly predicted as ``mildly defective'' and ``severely defective'', respectively. Therefore, TP$=174$ and FN $= 10$. For the ``functional'' class, FP is the number of observations that are incorrectly classified as ``functional'': 17 of the ``mildly defective'', 6 of  the``moderately defective'', and 24 of  the ``severely defective'' observations are incorrectly classified as ``functional'' so that FP $= 47$. For the ``functional'' class, TN is the number of observations that are correctly classified as not ``functional'', or the total number of observations excluding FP, TP and FN, $\text{TN} =310-47-174-10$. Note that TN is the sum of the remaining cells after deleting the row and column corresponding to ``functional'' in the confusion matrix in Figure~\ref{fig:cm_poly_rf_eg}. The metrics are computed analogously for each class.

Based on the definitions of TP, FP, TN, and FN, multiple metrics can be computed. For each class $j$, $j=1, \ldots, J$, the accuracy is calculated by,
\begin{equation*}
\Accuracy_j = \frac{\TP_j + \TN_j}{\TP_j + \TN_j+ \FP_j + \FN_j}, \quad j=1, \dots, J.
\end{equation*}
The overall accuracy for multi-classification is computed by taking the average of all accuracy values.

The precision and recall for class $j$ are given by,
\begin{equation*}
\Precision_j = \frac{\TP_j}{\TP_j + \FP_j}  \text{ and } \Recall_j = \frac{\TP_j}{\TP_j + \FN_j}, \quad j=1, \dots, J,
\end{equation*}
respectively. Precision measures the ratio of correctly predicted positive outcomes among all predicted positive observations, which reflects the accuracy of the minority classes. It is useful when the cost of the FP is very high or the cost of FN is low. Recall indicates how likely one observation is classified correctly among each class, and a high recall value indicates that the model is good at finding positive instances. The macro-precision average and macro-recall average are given by,
\begin{equation*}
\MacroAveragePrecision = J^{-1}\sum_{j=1}^J \Precision_j, \text{ and } \MacroAverageRecall = J^{-1}\sum_{j=1}^J \Recall_j,
\end{equation*}
respectively. Balanced accuracy (\BalancedAccuracy) is equal to \MacroAverageRecall, which averages the measurements across different classes. If the dataset is well balanced, \Accuracy~and \BalancedAccuracy~tend to be close. However, when there exists severe imbalance among class counts in the original dataset, they are quite different because each class contributes equally in \BalancedAccuracy, while \Accuracy~is heavily influenced by the majority class. Therefore,  \BalancedAccuracy~accounts for imbalanced problems when evaluating model performance.

The F1-score is the harmonic mean of precision and recall, providing a balanced measure of the model's ability to correctly classify both positive and negative instances. It is particularly useful in situations where there is a significant class imbalance, as it considers the performance of each class individually. Recall and precision have equal contributions to the F1-score calculation. In multi-classification problems, the F1-score for class $j$ is,
\begin{equation*}
\F_{1,j} = 2 \times \frac{\Precision_j \times \Recall_j}{\Precision_j+ \Recall_j}, \quad j=1, \dots, J.
\end{equation*}
The macro F1-score is defined as,
\begin{equation*}
\MacroF_1 = 2\times \frac{\MacroAveragePrecision \times \MacroAverageRecall}{\MacroAveragePrecision + \MacroAverageRecall}.
\end{equation*}
The weighted F1-score is defined as,
\begin{equation*}
\WeightedF_1 = 2 \times \sum_{j}\frac{N_j}{N_{\total}} \times \frac{\Precision_j \times \Recall_{j}}{\Precision_j+ \Recall_j},
\end{equation*}
where $N_j$ is the number of observations for class $j$ and $N_{\total}$ is the number of total observations. The weighted F1-score finds the average value of the F1-scores using the number of true observations for each class, while the Macro F1-score is the arithmetic mean of those F1-scores over classes. Compared with weighted F1-score value, Macro F1-score treats all classes equally regardless of class frequency, which considers the imbalanced data problem.

In binary classification, specificity is the ability of the model to correctly identify the negatives out of all actual negative cases, which is straightforward to evaluate the model in terms of performance of true negatives and false negatives. However, in multi-class problem settings, for each class considered as ``positive'', all other classes are collectively considered ``negative''. Calculating overall specificity among four classes requires averaging the specificity across all classes and considering the non-target classes as a single ``negative'' group, which will reduce specificity's ability to distinguish among each class. Instead, we introduce Matthew's Correlation Coefficient (MCC), which is a robust metric that incorporates all elements of the confusion matrix for the entire set of classifications. MCC is computed as:
\begin{equation*}
\MCC = \frac{m\times N_{\total} - \sum_{j}^J p_j \times t_j}{\sqrt{(N_{\total}^2- \sum_{j}^J p_{j}^2)(N_{\total}^2 - \sum_{j}^Jt_j^2)}}\,,
\end{equation*}
where $m$ is the total number of images correctly predicted, which is also the sum of $\TP$ and $\TN$ among all classes. As shown in Figure~\ref{fig:cm_poly_rf_eg}, $m$ is the sum of number of those diagonal elements in the confusion matrix, that is, $174+14+0+56 = 244$. $N_{\total}$ is the total number of images,  and $p_j$ is the number of images that class $j$ was predicted to have. In the above example, the values of $p_j$ are $221, 18, 0$ and $71$ for $j = 1, 2, 3, 4$. $t_j$ is the total number of images that class $j$ is truly predicted, which is the diagonal elements in the confusion matrix: 174, 14, 0, and 56 for $j = 1, 2, 3, 4$. MCC penalizes the model both for misclassifying minority class
instances and for incorrectly classifying non-instances as part of the minority class, providing
a balanced view of model accuracy that is robust to great performance on
the majority class alone, which is more reliable and informative than accuracy and F1-score
mentioned above. There are also other criteria such as the Cohen's Kappa and Hamming loss measures. Because Cohen's Kappa reflects similar information as MCC, we would not give details about it.

\subsection{EL Image Analysis Results}
\begin{figure}
	\begin{center}
		\begin{tabular}{cc}
			\includegraphics[width=.48\textwidth]{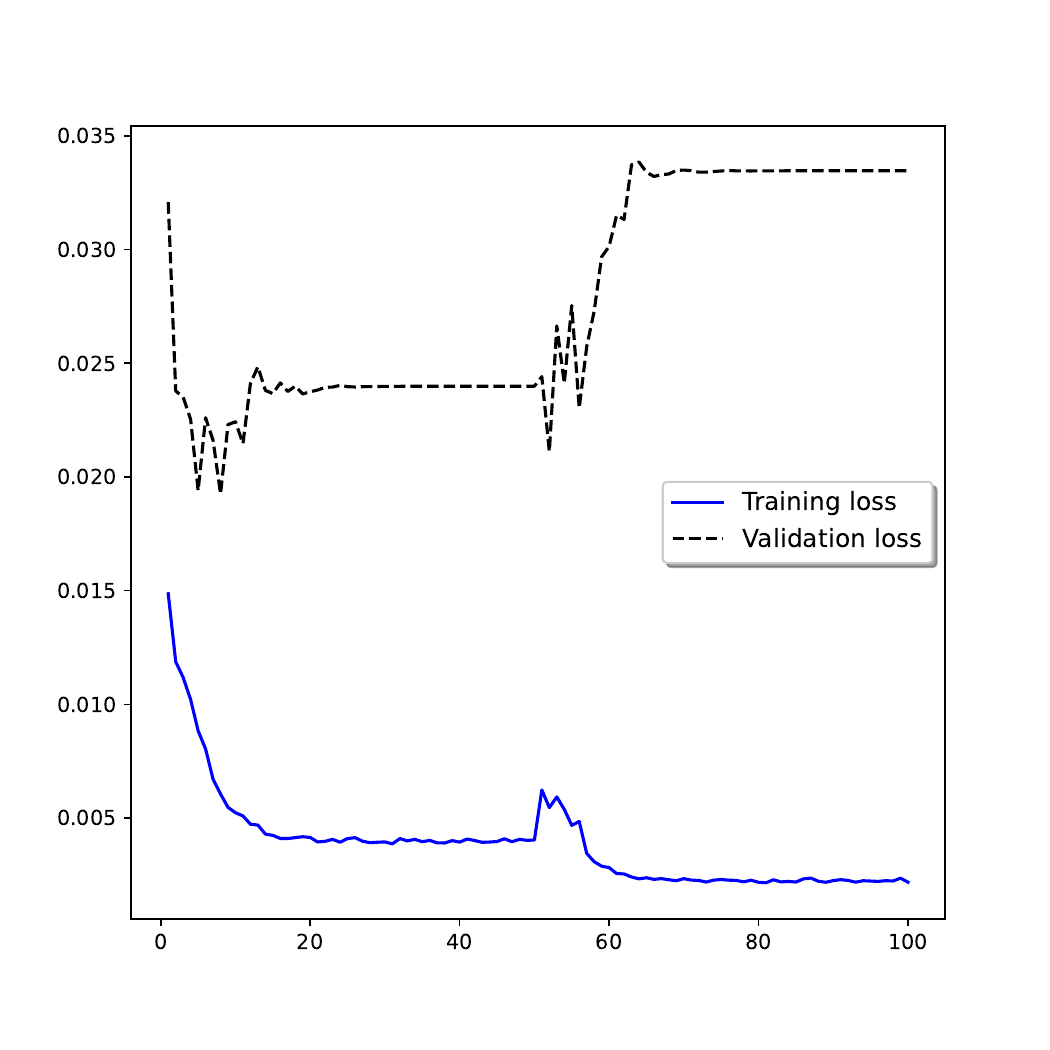} &
			\includegraphics[width=.48\textwidth]{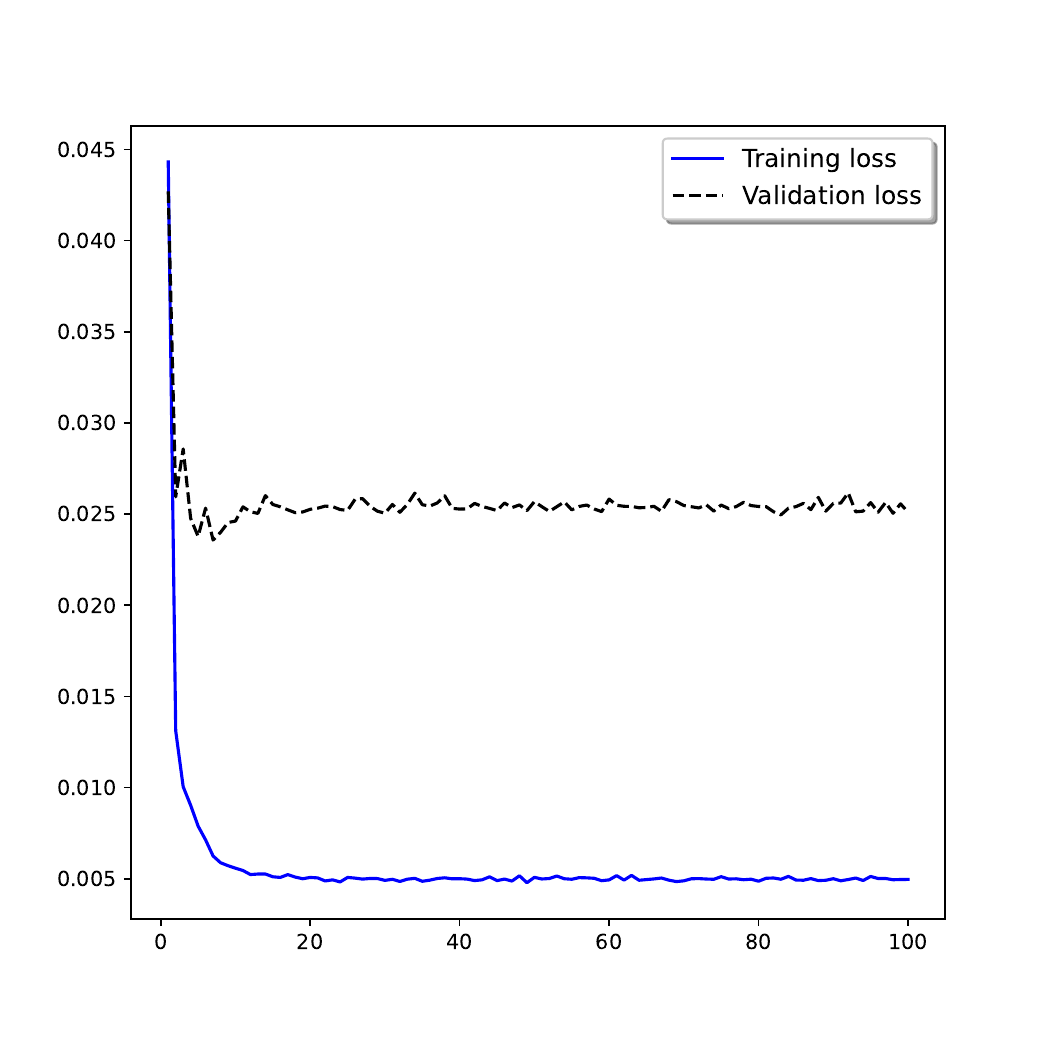}\\
			(a) VGG-19  & (b) ResNet-50\\
		\end{tabular}
		\caption{Loss of VGG-19 and ResNet-50 in polycrystalline PV modules.}
		\label{DL_loss}
	\end{center}
\end{figure}

Table~\ref{table: accmono.acc.poly} presents the accuracy of both ML and DL models in classifying PV modules. DL models perform much better than ML models in terms of prediction accuracy but require more intensive computation resources. Figure~\ref{DL_loss} shows the cross-entropy loss (\shortciteNP{CovTho}) of both the training and validation data in DL models. In the first stage, 50 epochs are trained with only the fully connected layers unfrozen, while in the second stage, all parameters are trained for 50 epochs. As we can see, the second stage reduces the loss and converges after 10 epochs. Slight spikes of loss in VGG-19 arise at the transition between the two stages because there is a brief period of adjustment where the loss experiences slight fluctuations as the model adapts to the new conditions.

\begin{figure}
	\begin{center}
		\begin{tabular}{cc}
			\includegraphics[width=.48\textwidth]{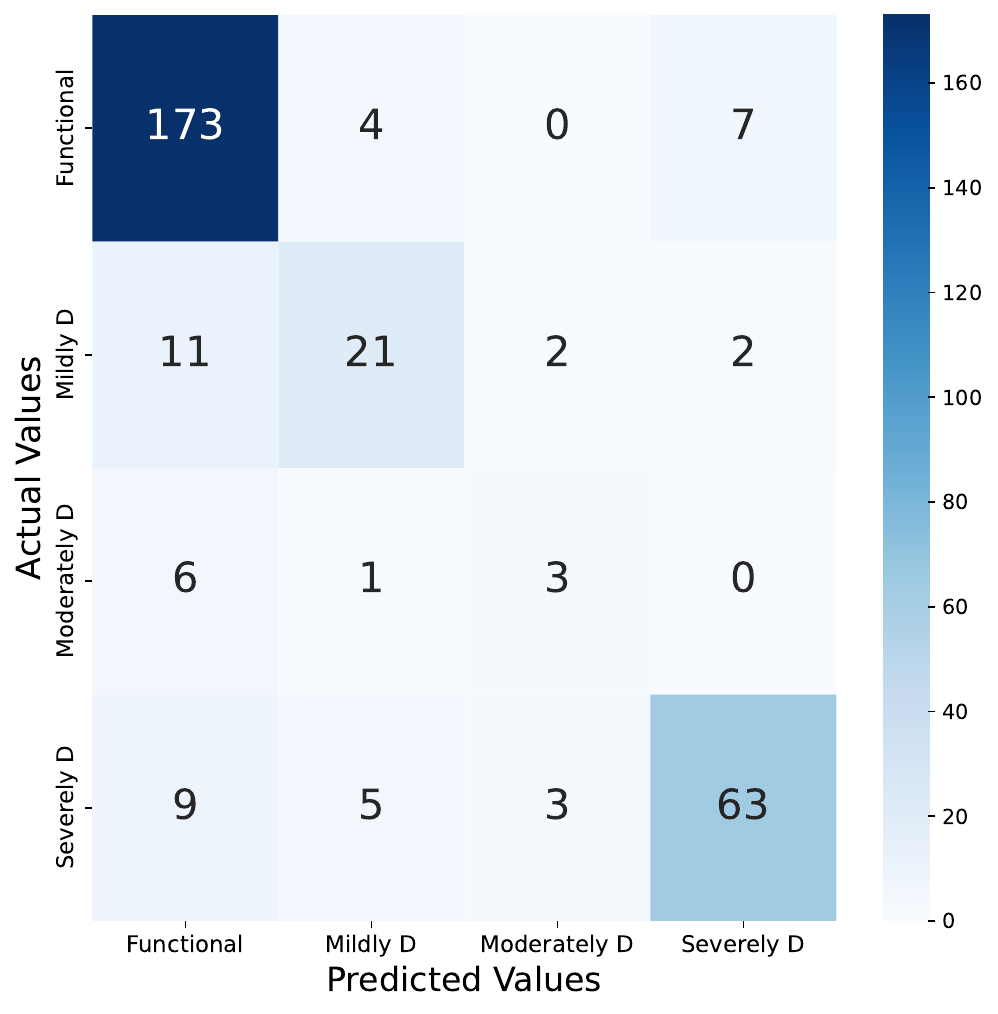} &
			\includegraphics[width=.48\textwidth]{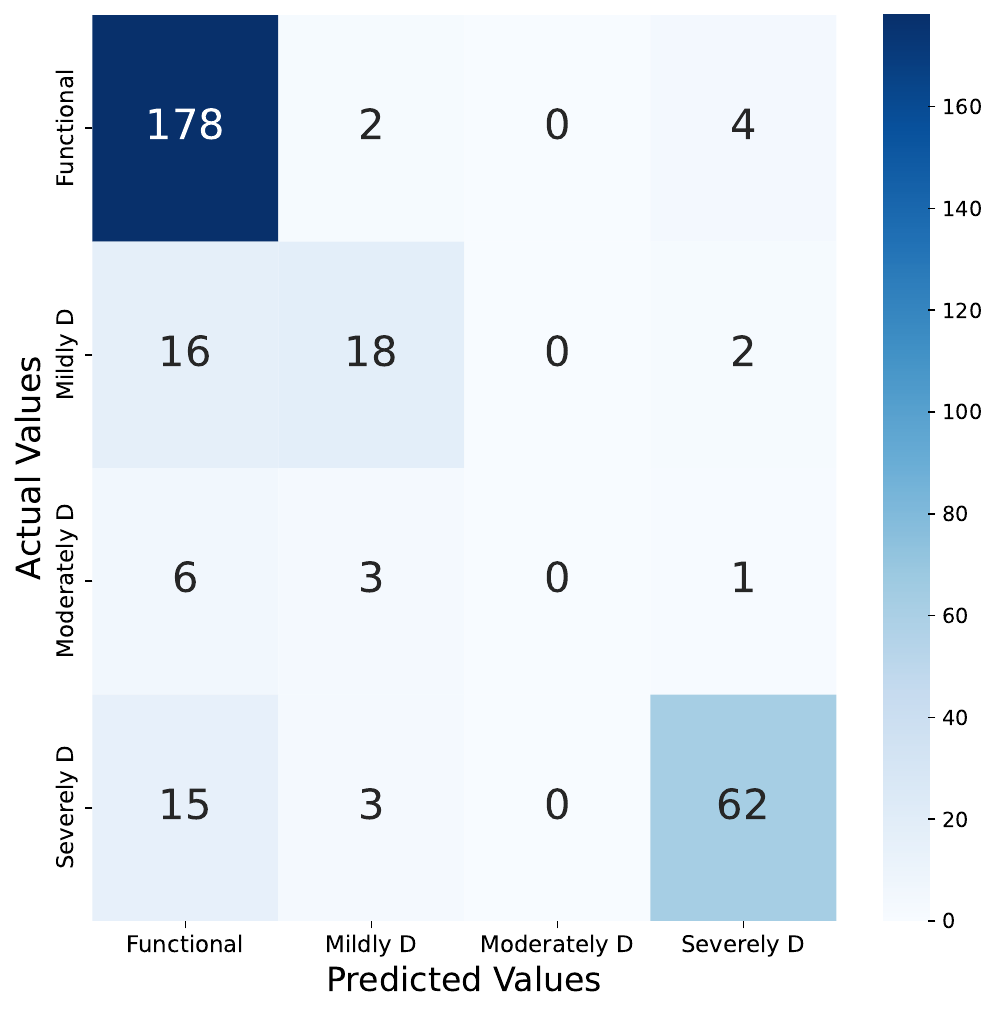}\\
			(a) VGG-19 & (b) ResNet-50\\
		\end{tabular}
		\caption{Confusion matrices of VGG-19 and ResNet-50 for polycrystalline PV modules.}
		\label{cm_DL}
	\end{center}
\end{figure}

The confusion matrices in Figure~\ref{cm_DL} show that both ML and DL models perform well, especially in the majority classes, which include the ``functional'' and ``severely defective'' panels, while for the ``mildly defective'' and ``moderately defective'' classes, they do not perform well. That is because there exists a severe imbalance problem in the dataset, which makes the predictions skewed in favor of the majority class and harms the model performance.

\begin{table}
	\caption{Accuracies of DL and ML models for monocrystalline and polycrystalline PV modules. Here, ``Logit'' stands for ``logistic regression''.}
	\centering
	\begin{tabular}{c |c c|  c c c}
		\hline\hline
\multirow{3}{*}{Monocrystalline}	 &	\multicolumn{2}{c|}{DL Models} & \multicolumn{3}{c}{ML Models} \\
		\cline{2-6}
		& VGG-19 & ResNet-50 & Logit & SVM & RF \\
		\cline{2-6}
		& 86.45$\%$ &  84.58$\%$ & 81.40 $\%$ & 83.26$\%$& 82.33$\%$  \\
		\hline\hline
\multirow{3}{*}{Polycrystalline} &
		\multicolumn{2}{c|}{DL Models} & \multicolumn{3}{c}{ML Models} \\
		\cline{2-6}
		& VGG-19 & ResNet-50 & Logit & SVM & RF \\
		\cline{2-6}
		& 87.75$\%$ &  86.13$\%$ & 80.00 $\%$ & 80.32$\%$& 86.13$\%$  \\
		\hline\hline
	\end{tabular}
	\label{table: accmono.acc.poly}
\end{table}

\subsubsection{Model Comparison}\label{model_comparison}

We compare both ML models (i.e., logistic regression, SVM, and RF) and DL models (i.e., the pre-trained VGG-19 and ResNet-50). The hyperparameter tuning part mainly lies in the specific parameter that relates to the model properties themselves. Each model is replicated 50 times with the same hyperparameter tuning procedure based on different random training and testing splits, and then the summary statistics including the median and standard deviation of each metric are calculated.

\begin{figure}
\begin{center}
\begin{tabular}{cc}
\includegraphics[width = 0.48\textwidth]{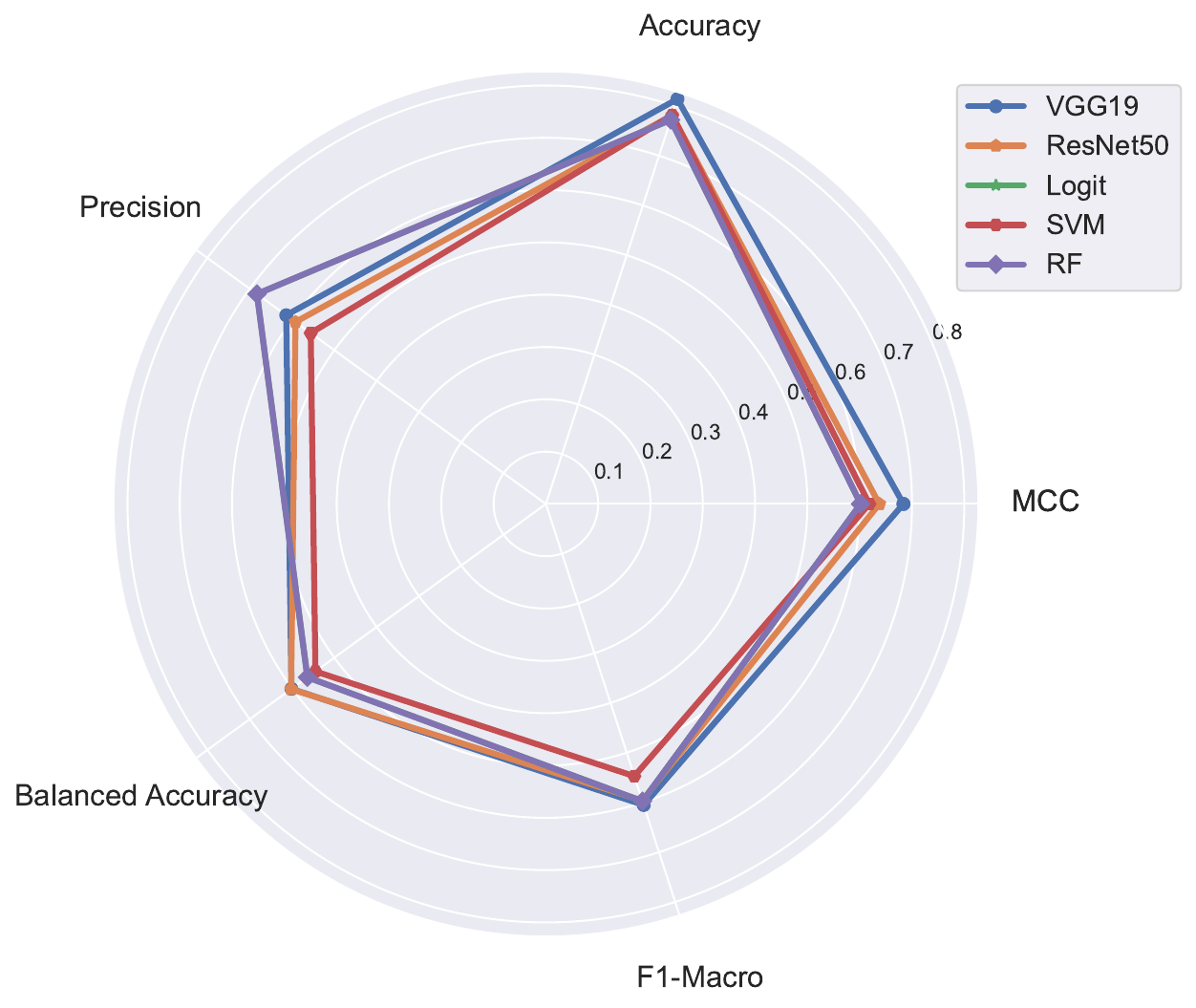} &
\includegraphics[width = 0.48\textwidth]{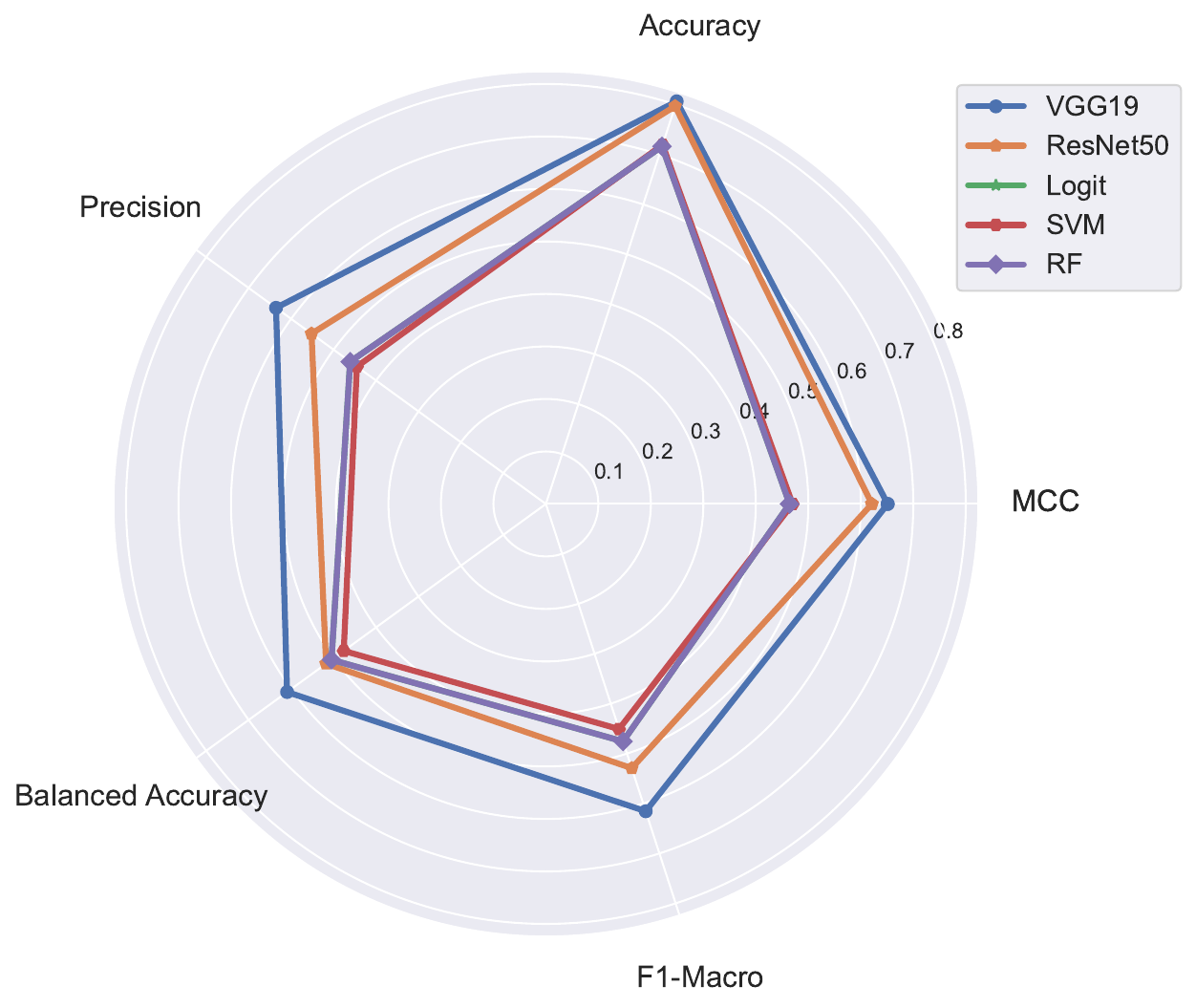}\\
(a) Monocrystalline & (b) Polycrystalline\\
\end{tabular}
\caption{Radar plots for the medians of metrics for various ML models among 50 replications for both the monocrystalline polycrystalline PV modules. Note the lines for the logistic model overlap with the lines for RF.} \label{fig:median.metric.radar.plot.mono.poly}
\end{center}
\end{figure}

Although in terms of accuracy, for both monocrystalline and polycrystalline PV modules, DL models perform better overall than ML models. Figure~\ref{fig:median.metric.radar.plot.mono.poly} gives radar plots that compare ML model performances with respect to the medians over all classes of the evaluation metrics in classifying the PV modules. Logistic regression and RF perform best in classifying monocrystalline PV modules. For polycrystalline PV modules, VGG-19 and ResNet-50, in this order, are the best taking all metrics into consideration. However, in terms of balanced accuracy, the ML models are almost as good as ResNet-50, which indicates that ML models also perform well in the minority class (``mildly defective'' and ``moderately defective'') predictions.

Regarding the stability and robustness of the model, for the monocrystalline PV modules as shown in Figure~\ref{fig:median.metric.radar.plot.mono.poly}(a), ResNet-50 exhibits smaller standard deviations among five models. Note that VGG-19 is most complex and requires intensive computation resources. In terms of stability, VGG-19 performs worse than ML models. Among all ML models under consideration, SVM typically has slightly lower standard deviations than logistic regression and RF, except when considering precision, and it is obvious that in terms of precision, DL models have more stable performance than ML models.

While for polycrystalline PV modules, the comparison results are not the same. Except for precision, ResNet-50 dominates other models with the smallest metric variations. Balanced accuracy is the metric with the largest standard deviation, and DL models especially ResNet-50 perform much better in terms of this metric. It is surprising that even with data augmentation and heavy computation, VCG-19 does not exhibit robustness, and performs worse than ML models trained on the original data in terms of balanced accuracy, MCC, and F1-macro score as shown in Figure~\ref{fig:median.metric.radar.plot.mono.poly}.

The behavior and performance of ML and DL models can vary depending on the specific challenges, dataset attributes, and implementation approaches. Although there indeed exists a tendency for smaller deviations in DL models, these observations are not universally applicable. It is crucial to evaluate and compare the model on a case-by-case basis.

\begin{figure}
\begin{center}
\begin{tabular}{cc}
\includegraphics[width = 0.48\textwidth]{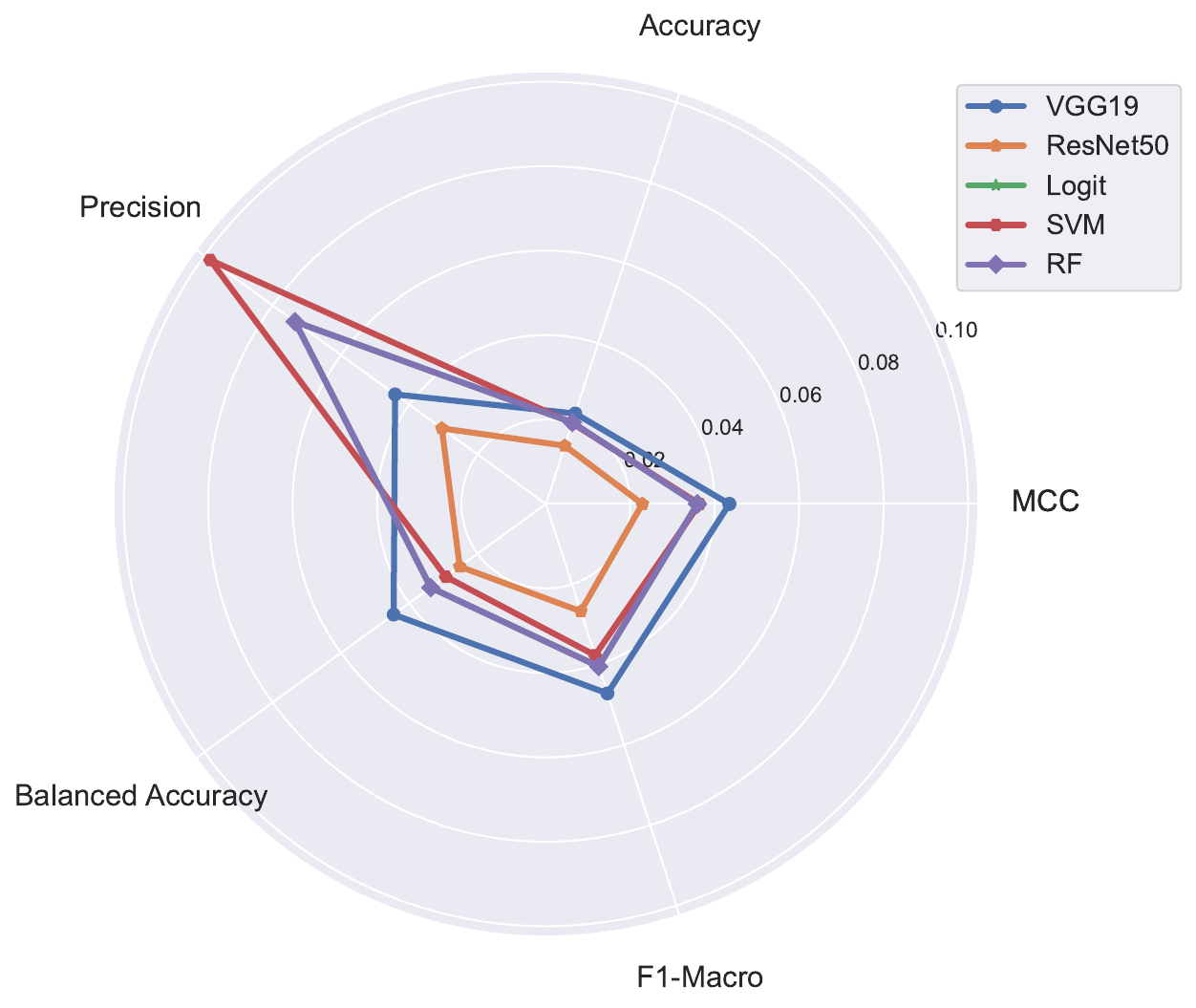} &
\includegraphics[width = 0.48\textwidth]{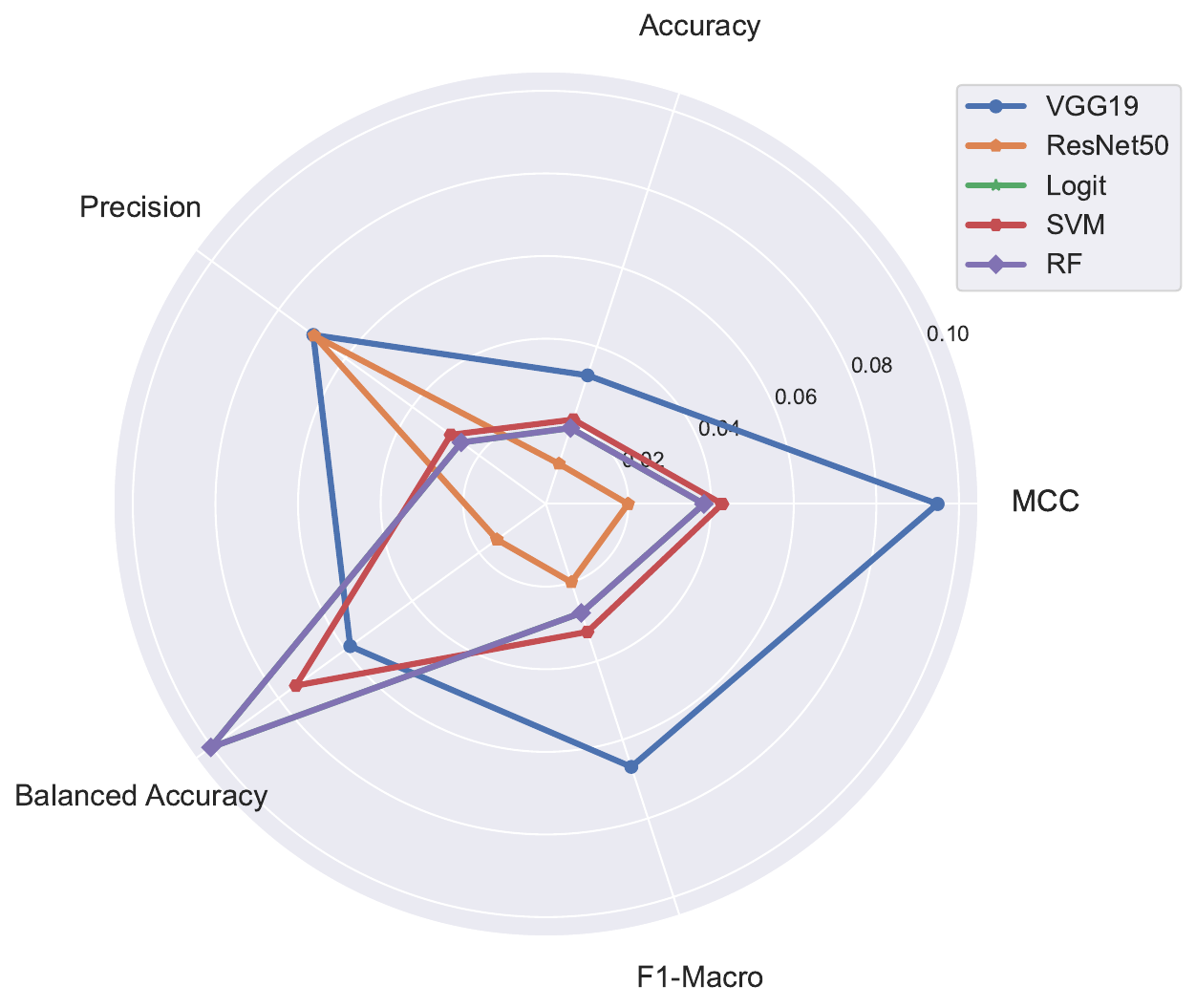}\\
(a) Monocrystalline & (b) Polycrystalline\\
\end{tabular}
\caption{Radar plots for the standard deviations of metrics for various ML models among 50 replications for both the monocrystalline polycrystalline PV modules. Note the lines for the logistic model overlap with the lines for RF.} \label{fig:std.metric.radar.plot.mono.poly}
\end{center}
\end{figure}

\subsubsection{Metrics Comparison}

Based on the study above, we can see that when using the same model and dataset, different metrics assess model performance from entirely different perspectives. With imbalanced data, accuracy often reflects the prediction of the majority class, which is most stable in both ML and DL models for both monocrystalline and polycrystalline PV modules among all metrics we explore. MCC offers a more nuanced view that accounts for the proportion of each type of correct and incorrect classifications, and also, the imbalanced problems. Besides, balanced accuracy takes into account the performance of the model across multiple classes by calculating the average recall measurements of each class. It provides a more comprehensive evaluation of the model's overall ability to classify correctly across all classes. Indeed, a model that achieves a high overall accuracy may not necessarily perform well in terms of balanced accuracy, particularly when it struggles with predicting the minority class. For instance, SVM achieves high accuracy in polycrystalline PV modules but it does not perform well in terms of balanced accuracy as shown in Figure~\ref{fig:median.metric.radar.plot.mono.poly}. Similarly, for monocrystalline PV modules, VGG-19 achieves the highest scores in terms of accuracy and MCC, but this does not hold true for balanced accuracy. The reason is that MCC's calculation incorporates all elements of the confusion matrix (TP, TN, FP, FN), which balances the positive and negative classes equally with consideration of all error types. On the other hand, the balanced accuracy considers performance across all classes by calculating an average, which makes it particularly sensitive to the performance of the minority class, therefore, it offers a more straightforward way for illustrating average accuracy per class.

\begin{figure}
	\begin{center}
			\includegraphics[width=.55\textwidth]{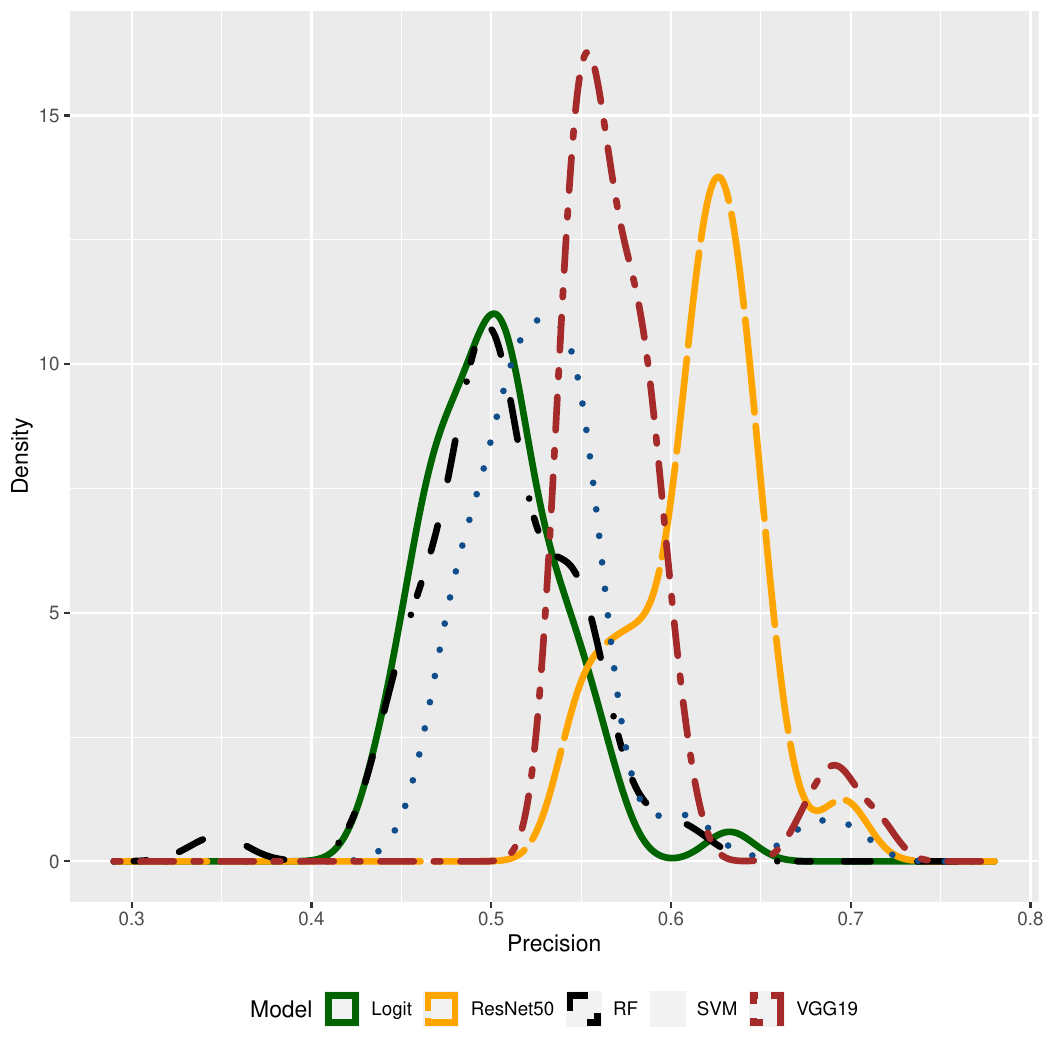}
		\caption{Density plot of precision metric in polycrystalline.}
		\label{density:precision}
	\end{center}
\end{figure}

Precision quantifies the model's ability to avoid false positive errors. It is interesting that in monocrystalline PV modules, precision is the metric with the largest variations among ML models, especially SVM. It is surprising that DL models have far larger precision standard deviation than ML models under polycrystalline PV modules. Figure~\ref{density:precision} gives a density plot of precision values for polycrystalline PV modules. However, precision alone might not provide a complete picture of model's performance, especially under scenarios of imbalanced data or situations where we attach equal importance to both false positives and false negatives, and it is often used in combination with other metrics such as recall and F1-score to obtain a comprehensive assessment of the model's predictive capabilities. The Macro F1-score calculates the F1-score independently for each class and then takes the average across all classes, giving equal weight to each class, which makes it a robust metric for evaluating classification model across imbalanced datasets. However, the robustness of F1-score can be influenced by certain factors such as the severity of imbalanced problems and the specific implementation of algorithms and assumption. For the EL image dataset, because data in monocrystalline and polycrystalline PV module is of similar severity of defectiveness, the main difference in model performance is most likely influenced by minority class predictions.

\section{Concluding Remarks and Suggestions for Future Research}\label{sec:conclusion.remark}

This paper presents a comprehensive comparison of traditional ML and modern DL models. We evaluate their performance in the classification of solar cell EL images from monocrystalline and polycrystalline sources. Various performance metrics are considered to provide a thorough analysis and comparison of these two methodologies. In terms of accuracy, DL models generally outperform ML models. However, when dealing with imbalanced issues, DL models do not necessarily outperform in metrics like balanced accuracy and F1-Macro score. For various PV module types, the effectiveness of DL and ML models differs. Moreover, after evaluating each model 50 times, the performance and behavior of ML and DL models are influenced by distinct challenges, dataset characteristics, and implementation strategies.

For metrics comparison, although both DL and ML models perform well in terms of accuracy which tends to favor the majority class, the result is not satisfactory for predicting minority classes that include ``mildly defective'' and ``moderately defective''. The severe imbalance problem in the original EL dataset poses challenges for ML and DL methods because they tend to be biased towards the majority class. They struggle to learn patterns and make accurate predictions for the minority class due to the lack of sufficient examples.  \shortciteN{robustimbalance2021} mentioned that the robustness of classification algorithms can be affected by imbalance class labels.
As a result, the models may have a high false negative rate for the minority class and lower overall performance. Also, as mentioned in \shortciteN{AIreliability2023}, imbalanced data is an important issue for AI reliability. Therefore, it is essential to measure the impact of chosen approach and evaluation metrics on specific classes.

We want to provide some more discussion on the inclusion of those ML and DL methods in this paper. Our selection of models is not solely based on complexity for the highest accuracy With more complex models and advanced GPUs, it is likely that the accuracy will be higher.
Given various kinds of models and evaluation metrics, even using the same dataset, various metrics may identify different models as the best one. In terms of model accuracy, given the same problem (four class label) and the same dataset, we found a few studies that have slightly better performance than ours. In particular, SeFNet combined with generative adversarial network (GAN) model for data augmentation (\shortciteNP{ZHAO2023126605}) is applied to deal with multi-class classification problems of defectiveness detection of solar cells based on EL images. Considering four label problems, their accuracy of monocrystalline PV modules can reach 87.5$\%$ and 81.79$\%$ for polycrystalline PV modules. However, they do not do replications and there exist randomness in the neural network structure. Besides, most of them applied GAN method for data augmentation, and it requires intensive computation resources. Another method, which uses deep feature-based (DFB) feature extraction combining with SVM model, can achieve an accuracy of 90.57$\%$ (\shortciteNP{DEMIRCI2021114810}). However, DFB model comes with numerous layers and parameters, require significant computational resources for both training and inference.

Although the overall model accuracy of those ML and DL methods in this paper is slightly lower than those models mentioned above, the difference (around 1\% $\sim$ 2\%) is not significant. Besides, based on pretrained model structure, we do not have to train the model from scratch, which saves much time and computations, and thus enables us to carry out the comprehensive model and metric comparisons. Also, 50 replications are implemented in our study for each model in both monocrystalline and polycrystalline scenarios to avoid randomness and guarantee the robustness of metrics.

In future work, we aim to address the imbalance problem so that models perform better in minority class predictions. We consider sampling methods (e.g., \shortciteNP{samplingAI2023}) to create a more balanced training set, modifications of algorithms including gradient boosting which helps adjust their internal calculations to give more weight to the minority class during training, cost-sensitive learning approaches (e.g., \shortciteNP{Khan2015CostSensitiveLO}) by penalizing misclassification of the minority class more heavily, representation extraction methods (e.g., \shortciteNP{guo2022tight}) to learn useful features from data for higher efficiency and accuracy, ensemble methods by combining multiple models trained on balanced subsets of the data to improve performance, and also trying more advanced data augmentation methods to increase observations from minority classes.

\section*{Data Availability Statement}
The EL image data and computing codes used in this paper are publicly available at GitHub repository: \url{https://github.com/yili-hong/ELImageClassifictionMLComparison}.

\section*{Acknowledgments}
The authors thank the editor, associate editor, and two referees, for their valuable comments that helped improve the paper significantly. The authors acknowledge the Advanced Research Computing program at Virginia Tech for providing computational resources.




\section*{About the Authors}

\hspace{1.5em}\textbf{Xinyi Song} is a PhD student of Statistics at Virginia Tech. Her research interest is on the interface between statistics and artificial intelligence.

\textbf{Kennedy Odongo} is Assistant Professor of Data Analytics at Hamline University. His research interests include data science, machine learning, and analytics.

\textbf{Francis G. Pascual} is Emeritus Associate Professor at Washington State University. His research interests include engineering reliability and quality control.

\textbf{Yili Hong} is Professor of Statistics at Virginia Tech. His research interests include engineering statistics, machine learning and statistical computing, and biostatistics. He is a Senior Member of ASQ. His email address is yilihong@vt.edu.

\section*{Disclosure Statement}

No potential conflict of interest was reported by the authors.

\section*{Funding}

Hong's work was partially supported by the Virginia Tech College of Science (COS) Research Equipment Fund and the COS Dean's Discovery Fund (Award: 452021).


\begin{thebibliography}{}

\bibitem[\protect\citeauthoryear{Al-Malla, Jafar, and Ghneim}{Al-Malla
  et~al.}{2022}]{AlMalla2022PretrainedCA}
Al-Malla, M.~A., A.~Jafar, and N.~Ghneim (2022).
\newblock Pre-trained {CNN}s as feature-extraction modules for image
  captioning.
\newblock {\em ELCVIA Electronic Letters on Computer Vision and Image
  Analysis\/}~{\em 21}, 1--16.

\bibitem[\protect\citeauthoryear{Bdour, Dalala, Al-Addous, Radaideh, and
  Al-Sadi}{Bdour et~al.}{2020}]{su12166416}
Bdour, M., Z.~Dalala, M.~Al-Addous, A.~Radaideh, and A.~Al-Sadi (2020).
\newblock A comprehensive evaluation on types of microcracks and possible
  effects on power degradation in photovoltaic solar panels.
\newblock {\em Sustainability\/}~{\em 12\/}(16), 6416.

\bibitem[\protect\citeauthoryear{Breiman}{Breiman}{2001}]{breiman2001random}
Breiman, L. (2001).
\newblock Random forests.
\newblock {\em Machine Learning\/}~{\em 45\/}(1), 5--32.

\bibitem[\protect\citeauthoryear{Buerhop, Deitsch, Maier, Gallwitz, Berger,
  Doll, Hauch, Camus, and Brabec}{Buerhop et~al.}{2018}]{Buerhop2018}
Buerhop, C., S.~Deitsch, A.~Maier, F.~Gallwitz, S.~Berger, B.~Doll, J.~Hauch,
  C.~Camus, and C.~J. Brabec (2018).
\newblock A benchmark for visual identification of defective solar cells in
  electroluminescence imagery.
\newblock In {\em European PV Solar Energy Conference and Exhibition (EU
  PVSEC)}, Volume~35, pp.\  1287--1289.

\bibitem[\protect\citeauthoryear{Cortes and Vapnik}{Cortes and
  Vapnik}{1995}]{SVM1995}
Cortes, C. and V.~Vapnik (1995).
\newblock Support-vector networks.
\newblock {\em Machine Learning\/}~{\em 20}, 273--297.

\bibitem[\protect\citeauthoryear{Cover and Thomas}{Cover and
  Thomas}{1991}]{CovTho}
Cover, T.~M. and J.~A. Thomas (1991).
\newblock {\em Elements of Information Theory}.
\newblock New York: Wiley.

\bibitem[\protect\citeauthoryear{Deitsch, Christlein, Berger, Buerhop-Lutz,
  Maier, Gallwitz, Maier, Gallwitz, and Riess}{Deitsch
  et~al.}{2019}]{SergiuDeitschEL2019}
Deitsch, S., V.~Christlein, S.~Berger, C.~Buerhop-Lutz, A.~Maier, F.~Gallwitz,
  A.~Maier, F.~Gallwitz, and C.~Riess (2019).
\newblock Automatic classification of defective photovoltaic module cells in
  electroluminescence images.
\newblock {\em Solar Energy\/}~{\em 185}, 455--468.

\bibitem[\protect\citeauthoryear{Deitsch, Christlein, Berger, Buerhop-Lutz,
  Maier, Gallwitz, and Riess}{Deitsch et~al.}{2019}]{Deitsch2019}
Deitsch, S., V.~Christlein, S.~Berger, C.~Buerhop-Lutz, A.~Maier, F.~Gallwitz,
  and C.~Riess (2019, June).
\newblock Automatic classification of defective photovoltaic module cells in
  electroluminescence images.
\newblock {\em Solar Energy\/}~{\em 185}, 455--468.

\bibitem[\protect\citeauthoryear{Demirci, Beşli, and
  G\"{u}m\"{u}sc\"{u}}{Demirci et~al.}{2021}]{DEMIRCI2021114810}
Demirci, M.~Y., N.~Beşli, and A.~G\"{u}m\"{u}sc\"{u} (2021).
\newblock Efficient deep feature extraction and classification for identifying
  defective photovoltaic module cells in electroluminescence images.
\newblock {\em Expert Systems with Applications\/}~{\em 175}, 114810.

\bibitem[\protect\citeauthoryear{Deng, Dong, Socher, Li, Li, and Li}{Deng
  et~al.}{2009}]{ImageNet2009}
Deng, J., W.~Dong, R.~Socher, L.~Li, K.~Li, and F.~Li (2009).
\newblock {ImageNet}: A large-scale hierarchical image database.
\newblock In {\em 2009 IEEE Conference on Computer Vision and Pattern
  Recognition({CVPR})}, pp.\  248--255.

\bibitem[\protect\citeauthoryear{Fuyuki, Kondo, Yamazaki, Takahashi, and
  Uraoka}{Fuyuki et~al.}{2005}]{wave11502005}
Fuyuki, T., H.~Kondo, T.~Yamazaki, Y.~Takahashi, and Y.~Uraoka (2005).
\newblock Photographic surveying of minority carrier diffusion length in
  polycrystalline silicon solar cells by electroluminescence.
\newblock {\em Applied Physics Letters\/}~{\em 86\/}(26), 262108.

\bibitem[\protect\citeauthoryear{Goodfellow, Bengio, and Courville}{Goodfellow
  et~al.}{2016}]{Goodfellow2016}
Goodfellow, I., Y.~Bengio, and A.~Courville (2016).
\newblock {\em Deep Learning}.
\newblock MIT Press.

\bibitem[\protect\citeauthoryear{Grandini, Bagli, and Visani}{Grandini
  et~al.}{2020}]{Grandini2020MetricsFM}
Grandini, M., E.~Bagli, and G.~Visani (2020).
\newblock Metrics for multi-class classification: an overview.
\newblock {\em ArXiv\/}~{\em abs/2008.05756}.

\bibitem[\protect\citeauthoryear{Guo, Chen, Wang, Yang, Deng, Huang, Carin, Li,
  and Tao}{Guo et~al.}{2022}]{guo2022tight}
Guo, Q., J.~Chen, D.~Wang, Y.~Yang, X.~Deng, J.~Huang, L.~Carin, F.~Li, and
  C.~Tao (2022).
\newblock Tight mutual information estimation with contrastive Fenchel-Legendre
  optimization.
\newblock {\em Advances in Neural Information Processing Systems\/}~{\em 35},
  28319--28334.

\bibitem[\protect\citeauthoryear{Hamada and Nelder}{Hamada and
  Nelder}{1997}]{GLMMJ1997}
Hamada, M. and J.~A. Nelder (1997).
\newblock Generalized linear models for quality-improvement experiments.
\newblock {\em Journal of Quality Technology\/}~{\em 29\/}(3), 292--304.

\bibitem[\protect\citeauthoryear{He, Zhang, Ren, and Sun}{He
  et~al.}{2015}]{resnet2016he}
He, K., X.~Zhang, S.~Ren, and J.~Sun (2015).
\newblock Deep residual learning for image recognition.
\newblock In {\em 2016 IEEE Conference on Computer Vision and Pattern
  Recognition ({CVPR})}, pp.\  770--778.

\bibitem[\protect\citeauthoryear{Hong, Lian, Xu, Min, Wang, Freeman, and
  Deng}{Hong et~al.}{2023}]{AIreliability2023}
Hong, Y., J.~Lian, L.~Xu, J.~Min, Y.~Wang, L.~J. Freeman, and X.~Deng (2023).
\newblock Statistical perspectives on reliability of artificial intelligence
  systems.
\newblock {\em Quality Engineering\/}~{\em 35\/}(1), 56--78.

\bibitem[\protect\citeauthoryear{Iqbal}{Iqbal}{2018}]{haris}
Iqbal, H. (2018).
\newblock Plotneuralnet v1.0.0, https://github.com/harisiqbal88/plotneuralnet.

\bibitem[\protect\citeauthoryear{Khan, Hayat, Bennamoun, Sohel, and
  Togneri}{Khan et~al.}{2015}]{Khan2015CostSensitiveLO}
Khan, S.~H., M.~Hayat, Bennamoun, F.~Sohel, and R.~B. Togneri (2015).
\newblock Cost-sensitive learning of deep feature representations from
  imbalanced data.
\newblock {\em IEEE Transactions on Neural Networks and Learning
  Systems\/}~{\em 29}, 3573--3587.

\bibitem[\protect\citeauthoryear{K\"{o}ntges, Kurtz, Packard, Jahn, Berger,
  Kato, Friesen, Liu, and Iseghem}{K\"{o}ntges
  et~al.}{2014}]{FailureCriterion2014}
K\"{o}ntges, M., S.~Kurtz, C.~Packard, U.~Jahn, K.~A. Berger, K.~Kato,
  T.~Friesen, H.~Liu, and M.~V. Iseghem (2014).
\newblock Review of failures of photovoltaic modules, technical report.

\bibitem[\protect\citeauthoryear{Krizhevsky, Sutskever, and Hinton}{Krizhevsky
  et~al.}{2017}]{Alexnet2017}
Krizhevsky, A., I.~Sutskever, and G.~E. Hinton (2017).
\newblock Imagenet classification with deep convolutional neural networks.
\newblock {\em Communications of the ACM\/}~{\em 60\/}(6), 84--90.

\bibitem[\protect\citeauthoryear{Lian, Freeman, Hong, and Deng}{Lian
  et~al.}{2021}]{robustimbalance2021}
Lian, J., L.~Freeman, Y.~Hong, and X.~Deng (2021).
\newblock Robustness with respect to class imbalance in artificial intelligence
  classification algorithms.
\newblock {\em Journal of Quality Technology\/}~{\em 53\/}(5), 505--525.

\bibitem[\protect\citeauthoryear{Lingras and Butz}{Lingras and
  Butz}{2007}]{LINGRAS20073782}
Lingras, P. and C.~Butz (2007).
\newblock Rough set based 1-v-1 and 1-v-r approaches to support vector machine
  multi-classification.
\newblock {\em Information Sciences\/}~{\em 177\/}(18), 3782--3798.

\bibitem[\protect\citeauthoryear{Monti, Tootoonian, and Cao}{Monti
  et~al.}{2018}]{AD2018}
Monti, R.~P., S.~Tootoonian, and R.~Cao (2018).
\newblock Avoiding degradation in deep feed-forward networks by phasing out
  skip-connections.
\newblock In V.~K{\r{u}}rkov{\'a}, Y.~Manolopoulos, B.~Hammer, L.~Iliadis, and
  I.~Maglogiannis (Eds.), {\em Artificial Neural Networks and Machine Learning
  -- ICANN 2018}, Cham, pp.\  447--456. Springer International Publishing.

\bibitem[\protect\citeauthoryear{Rahman and Chen}{Rahman and
  Chen}{2020}]{MRUDefectinspec2020}
Rahman, M. R.~U. and H.~Chen (2020).
\newblock Defects inspection in polycrystalline solar cells electroluminescence
  images using deep learning.
\newblock {\em IEEE Access\/}~{\em 8}, 40547--40558.

\bibitem[\protect\citeauthoryear{Rodriguez, Gonzalez, Fernandez, Rodriguez,
  Delgado, and Bellman}{Rodriguez et~al.}{2021}]{ASCDRA}
Rodriguez, A., C.~Gonzalez, A.~Fernandez, F.~Rodriguez, T.~Delgado, and
  M.~Bellman (2021).
\newblock Automatic solar cell diagnosis and treatment.
\newblock {\em Journal of Intelligent Manufacturing\/}~{\em 32}, 1163--1172.

\bibitem[\protect\citeauthoryear{Rokach and Maimon}{Rokach and
  Maimon}{2005}]{Rokach2005}
Rokach, L. and O.~Maimon (2005).
\newblock {\em Decision Trees}.
\newblock Boston, MA: Springer US.

\bibitem[\protect\citeauthoryear{Shaha and Pawar}{Shaha and
  Pawar}{2018}]{8474802}
Shaha, M. and M.~Pawar (2018).
\newblock Transfer learning for image classification.
\newblock In {\em 2018 Second International Conference on Electronics,
  Communication and Aerospace Technology (ICECA)}, pp.\  656--660.

\bibitem[\protect\citeauthoryear{Simonyan and Zisserman}{Simonyan and
  Zisserman}{2014}]{vgg16}
Simonyan, K. and A.~Zisserman (2014).
\newblock Very deep convolutional networks for large-scale image recognition.
\newblock {\em arXiv:1409.1556\/}.

\bibitem[\protect\citeauthoryear{Sovetkin, Achterberg, Weber, and
  Pieters}{Sovetkin et~al.}{2020}]{Sovetkin2020EncoderDecoderSS}
Sovetkin, E., E.~J. Achterberg, T.~Weber, and B.~E. Pieters (2020).
\newblock Encoder-decoder semantic segmentation models for electroluminescence
  images of thin-film photovoltaic modules.
\newblock {\em IEEE Journal of Photovoltaics\/}~{\em 11}, 444--452.

\bibitem[\protect\citeauthoryear{Szegedy, Liu, Jia, Sermanet, Reed, Anguelov,
  Erhan, Vanhoucke, and Rabinovich}{Szegedy et~al.}{2015}]{7298594}
Szegedy, C., W.~Liu, Y.~Jia, P.~Sermanet, S.~Reed, D.~Anguelov, D.~Erhan,
  V.~Vanhoucke, and A.~Rabinovich (2015).
\newblock Going deeper with convolutions.
\newblock In {\em 2015 IEEE Conference on Computer Vision and Pattern
  Recognition (CVPR)}, pp.\  1--9.

\bibitem[\protect\citeauthoryear{Taşçioğlu, Taşkın, and
  Vardar}{Taşçioğlu et~al.}{2016}]{PVshareSM2016}
Taşçioğlu, A., O.~Taşkın, and A.~Vardar (2016).
\newblock A power case study for monocrystalline and polycrystalline solar
  panels in bursa city, turkey.
\newblock {\em International Journal of Photoenergy\/}~{\em 2016}, 1--7.

\bibitem[\protect\citeauthoryear{Tan, Liao, Bai, Deng, Zhao, and Zhao}{Tan
  et~al.}{2019}]{DenoisingCNN}
Tan, Y., K.~Liao, X.~Bai, C.~Deng, Z.~Zhao, and B.~Zhao (2019).
\newblock Denoising convolutional neural networks based dust accumulation
  status evaluation of photovoltaic panel.
\newblock {\em IEEE International Conference on Energy Internet ({ICEI})\/},
  560--566.

\bibitem[\protect\citeauthoryear{Tang, Yang, Xiong, and Yan}{Tang
  et~al.}{2020}]{TANG2020453}
Tang, W., Q.~Yang, K.~Xiong, and W.~Yan (2020).
\newblock Deep learning based automatic defect identification of photovoltaic
  module using electroluminescence images.
\newblock {\em Solar Energy\/}~{\em 201}, 453--460.

\bibitem[\protect\citeauthoryear{Wang, Wang, and He}{Wang
  et~al.}{2023}]{samplingAI2023}
Wang, G., J.~Wang, and K.~He (2023).
\newblock Majority-to-minority resampling for boosting-based classification
  under imbalanced data.
\newblock {\em Applied Intelligence\/}~{\em 53\/}(4), 4541--4562.

\bibitem[\protect\citeauthoryear{Wang, Yang, and Yan}{Wang
  et~al.}{2019}]{WQ994713}
Wang, W., Q.~Yang, and W.~Yan (2019).
\newblock Deep learning based model for defect detection of mono-crystalline-si
  solar pv module cells in electroluminescence images using data augmentation.
\newblock In {\em 2019 IEEE PES Asia-Pacific Power and Energy Engineering
  Conference (APPEEC)}, pp.\  1--5.

\bibitem[\protect\citeauthoryear{Zhao, Song, Zhang, Sun, and Zhao}{Zhao
  et~al.}{2023}]{ZHAO2023126605}
Zhao, X., C.~Song, H.~Zhang, X.~Sun, and J.~Zhao (2023).
\newblock Hrnet-based automatic identification of photovoltaic module defects
  using electroluminescence images.
\newblock {\em Energy\/}~{\em 267}, 126605.

\end{thebibliography}

\end{document}